\documentstyle[12pt]{article}
\begin{document}

\begin{flushright}
SLAC--PUB--7473
\end{flushright}

\bigskip\bigskip
\centerline{\bf OPTIMAL RENORMALIZATION SCALE AND SCHEME FOR}
\vspace{8pt}
\centerline{{\bf EXCLUSIVE PROCESSES}\footnote{\baselineskip=14pt
Research supported in part by the U.S. Department of Energy and
National Science Foundation.}}
\vspace{22pt}
  \centerline{Stanley J. Brodsky}
\vspace{13pt}
  \centerline{\it Stanford Linear Accelerator Center}
  \centerline{\it Stanford University, Stanford, CA 94309}
\vspace*{0.5cm}
  \centerline{Chueng-Ryong Ji}
\vspace{13pt}
  \centerline{\it Department of Physics, North Carolina State University}
  \centerline{\it Raleigh, NC 27695}
\vspace*{0.5cm}
  \centerline{Alex Pang}
\vspace{13pt}
  \centerline{\it Nuclear Theory Center, Indiana University}
  \centerline{\it Bloomington, IN 47404}
\vspace*{0.5cm}
  \centerline{David G. Robertson}
\vspace{13pt}
  \centerline{\it Department of Physics, The Ohio State University}
  \centerline{\it Columbus, OH 43210}

\vfill
\pagebreak
\vspace*{0.9cm}
\begin{center} {\bf ABSTRACT} \end{center}
\noindent
We use the BLM method to fix the renormalization scale of the QCD
coupling in exclusive hadronic amplitudes such as the pion form factor
and the photon-to-pion transition form factor at large momentum
transfer.  Renormalization-scheme-independent commensurate scale
relations are established which connect the hard scattering subprocess
amplitudes that control exclusive processes to other QCD observables
such as the heavy quark potential and the electron-positron
annihilation cross section. The commensurate scale relation connecting
the heavy quark potential, as determined from lattice gauge theory, to
the photon-to-pion transition form factor is in excellent agreement
with $\gamma e \to \pi^0 e$ data assuming that the pion distribution
amplitude is close to its asymptotic form $\sqrt{3}f_\pi x(1-x)$.  We
also reproduce the scaling and normalization of the
$\gamma\gamma\rightarrow\pi^+\pi^-$ data at large momentum transfer.
Because the renormalization scale is small, we argue that the
effective coupling is nearly constant, thus accounting for the nominal
scaling behavior of the data.  However, the normalization of the
space-like pion form factor $F_\pi(Q^2)$ obtained from
electroproduction experiments is somewhat higher than that predicted
by the corresponding commensurate scale relation. This discrepancy may
be due to systematic errors introduced by the extrapolation of the
$\gamma^* p \to \pi^+ n$ electroproduction data to the pion pole.

\vfill
\baselineskip 17pt
\pagebreak

\section{Introduction}

One of the most critical problems in making reliable predictions in
quantum chromodynamics is how to deal with the dependence of the
truncated perturbative series on the choice of renormalization scale
$\mu$ and scheme for the QCD coupling $\alpha_s(\mu)$
\cite{Stevenson,Grunberg,BLM}. For processes such as jet production in
$e^+ e^-$ annihilation and heavy flavor production in hadron
collision, where only the leading and next-to-leading predictions are
known, the theoretical uncertainties from the choice of
renormalization scale and scheme are larger than the experimental
uncertainties. The ambiguities due to the renormalization conventions
are compounded in processes involving more than one physical scale.

Perturbative QCD has been used to analyze a number of exclusive
processes involving large momentum transfers, including the decay of
heavy hadrons to specific channels such as $B \to \pi \pi$ and
$\Upsilon \to p \bar p$, baryon form factors at large $t$, and fixed
$\theta_{c.m.}$ hadronic scattering amplitudes such as $\gamma p \to
\pi^+ n$ at high energies. As in the case of inclusive reactions,
factorization theorems for exclusive processes
\cite{BrodskyLepage,EfremovRad} allow the analytic separation of the
perturbatively-calculable short-distance contributions from the
long-distance non-perturbative dynamics associated with hadronic
binding.  For reviews of this formalism with many additional
references, see \cite{BLReview,StermanStoler}.

The scale ambiguities for the underlying quark-gluon subprocesses are
particularly acute in the case of QCD predictions for exclusive
processes, since the running coupling $\alpha_s$ enters at a high
power. Furthermore, since each external momentum entering an exclusive
reaction is partitioned among the many propagators of the underlying
hard-scattering amplitude, the physical scales that control these
processes are inevitably much softer than the overall momentum
transfer. Exclusive process phenomenology is further complicated by
the fact that the scales of the running couplings in the
hard-scattering amplitude depend themselves on the shape of the
hadronic wavefunctions.

The renormalization scale ambiguity problem can be resolved if one can
optimize the choices of scale and scheme according to some sensible
criteria. In the BLM procedure, the renormalization scales are chosen
such that all vacuum polarization effects from the QCD $\beta$
function are re-summed into the running couplings.  The coefficients
of the perturbative series are thus identical to the perturbative
coefficients of the corresponding conformally invariant theory with
$\beta=0.$ The BLM method has the important advantage of
``pre-summing" the large and strongly divergent terms in the PQCD
series which grow as $n!  (\alpha_s \beta_0 )^n$, {\em i.e.}, the
infrared renormalons associated with coupling constant renormalization
\cite{Mueller,BallBenekeBraun}. Furthermore, the renormalization
scales $Q^*$ in the BLM method are physical in the sense that they
reflect the mean virtuality of the gluon propagators
\cite{BLM,BallBenekeBraun,LepageMackenzie,Neubert}.  In fact, in the
$\alpha_V(Q)$ scheme, where the QCD coupling is defined from the heavy
quark potential, the renormalization scale is by definition the
momentum transfer caused by the gluon.

In this paper we will use the BLM method to fix the renormalization
scale of the QCD coupling in exclusive hadronic amplitudes such as the
pion form factor, the photon-to-pion transition form factor and
$\gamma \gamma \rightarrow \pi^+ \pi^-$ at large momentum transfer.
Renormalization-scheme-independent commensurate scale relations will
be established which connect the hard scattering subprocess amplitudes
that control these exclusive processes to other QCD observables such
as the heavy quark potential and the electron-positron annihilation
cross section.  Because the renormalization scale is small, we will
argue that the effective coupling is nearly constant, thus accounting
for the nominal scaling behavior of the data
\cite{JiSillLombard,JiAmiri}.

\section{Renormalization Scale Fixing in Exclusive Processes}

A basic principle of renormalization theory is the requirement that
the relations between physical observables must be independent of
renormalization scale and scheme conventions to any fixed order of
perturbation theory \cite{BoguluibovShirkov}. This property can be
explicitly expressed in the form of ``commensurate scale relations"
(CSR) \cite{CSR}. A primary example of a CSR is the generalized
Crewther relation \cite{CSR,BrodskyKataevGabaladzeLu}, in which the
radiative corrections to the Bjorken sum rule for deep inelastic
lepton-proton scattering at a given momentum transfer $Q$ are
predicted from measurements of the $e^+ e^-$ annihilation cross
section at a corresponding commensurate energy scale $\sqrt s \propto
Q$.

A scale-fixed relation between any two physical observables $A$ and
$B$ can be derived by applying BLM scale-fixing to their respective
perturbative predictions in, say, the $\overline {MS}$ scheme and then
algebraically eliminating $\alpha_{\overline {MS}}$. The choice of the
BLM scale ensures that the resulting CSR between $A$ and $B$ is
independent of the choice of the intermediate renormalization scheme
\cite{CSR}.  Thus, using this formalism one can relate any
perturbatively calculable observables, such as the annihilation ratio
$R_{e^+ e^-}$, the heavy quark potential, and the radiative
corrections to structure function sum rules, to each other without any
renormalization scale or scheme ambiguity \cite{BrodskyLu}.

The heavy-quark potential $V(Q^2)$ can be identified via the
two-particle-irreducible scattering amplitude of test charges, {\em
i.e.}, the scattering of an infinitely-heavy quark and antiquark at
momentum transfer $t = -Q^2.$ The relation
\begin{equation}
V(Q^2) = -  {4 \pi C_F \alpha_V(Q^2)\over Q^2},
\end{equation}
with $C_F=(N_C^2-1)/2 N_C=4/3$, then defines the effective charge
$\alpha_V(Q)$.  This coupling provides a physically-based alternative
to the usual ${\overline {MS}}$ scheme.  Recent lattice calculations
have provided strong constraints on the normalization and shape of
$\alpha_V(Q^2)$.

As in the corresponding case of Abelian QED, the scale $Q$ of the
coupling $\alpha_V(Q)$ is identified with the exchanged momentum.  All
vacuum polarization corrections due to fermion pairs are incorporated
in the usual vacuum polarization kernels defined in terms of physical
mass thresholds.  The first two terms $\beta_0 = 11 - 2 n_f/3$ and
$\beta_1 = 102 - 38n_f/3$ in the expansion of the $\beta$ function
defined from the logarithmic derivative of $\alpha_V(Q)$ are
universal, {\em i.e.}, identical for all effective charges at $Q^2 \gg
4m_f^2$.  The coefficient $\beta_2$ for $\alpha_V$ has recently been
calculated in the $\overline {MS}$ scheme \cite{MarkusPeter}.

The scale-fixed relation between $\alpha_V$ and the conventional
$\overline {MS}$ coupling is
\begin{equation}
\alpha_V(Q) = \alpha_{\overline {MS}}(e^{-5/6} Q) \left(1 -
\frac{2C_A}{3} {\alpha_{\overline {MS}}\over\pi} + \cdots\right),
\label{alpmsbar}
\end {equation}
above or below any quark mass threshold.  The factor $e^{-5/6} \simeq
0.4346$ is the ratio of commensurate scales between the two schemes to
this order.  It arises because of the conventions used in defining the
modified minimal subtraction scheme. The scale in the $\overline {MS}$
scheme is thus a factor $\sim 0.4$ smaller than the physical scale.
The coefficient $2C_A/3$ in the NLO term is a feature of the
non-Abelian couplings of QCD; the same coefficient would occur even if
the theory were conformally invariant with $\beta_0=0.$

As we shall see, the coupling $\alpha_V$ provides a natural scheme for
computing exclusive amplitudes. Once we relate, {\it e.g.}, form
factors to effective charges based on observables, there are no
ambiguities due to scale or scheme conventions.

The use of $\alpha_V$ as the expansion parameter with BLM scale-fixing
has also been found to be valuable in lattice gauge theory, greatly
increasing the convergence of perturbative expansions relative to
those using the bare lattice coupling \cite{LepageMackenzie}.  In
fact, new lattice calculations of the $\Upsilon$ spectrum
\cite{Davies} have been used to determine the normalization of the
static heavy quark potential and its effective charge:
\begin{equation}
\alpha_V^{(3)}(8.2~{\rm GeV}) = 0.196(3),
\label{alpv8.2}
\end{equation}
where the effective number of light flavors is $n_f = 3$. The
corresponding modified minimal subtraction coupling evolved to the $Z$
mass using Eq. (\ref{alpmsbar}) is given by
\begin{equation}
\alpha_{\overline{MS}}^{(5)}(M_Z) = 0.115(2).
\label{alpmsbarmz}
\end{equation}
This value is consistent with the world average of 0.117(5), but is
significantly more precise. These results are valid up to NLO.

Exclusive processes are particularly challenging to compute in quantum
chromodynamics because of their sensitivity to the unknown
non-perturbative bound state dynamics of the hadrons. However, in some
important cases, the leading power-law behavior of an exclusive
amplitude at large momentum transfer can be computed rigorously via a
factorization theorem which separates the soft and hard dynamics. For
example, the leading $1/Q^2$ fall-off of the meson form factors can be
computed as a perturbative expansion in the QCD coupling
\cite{BrodskyLepage,EfremovRad}:
\begin{equation}
F_M(Q^2)=\int^1_0 dx \int^1_0 dy
\phi_M(x, {\tilde Q}) T_H(x,y,Q^2)
\phi_M(y,{\tilde Q}),
\end{equation}
where $\phi_M(x,{\tilde Q})$ is the process-independent meson
distribution amplitude, which encodes the non-perturbative dynamics of
the bound valence Fock state up to the resolution scale ${\tilde Q}$,
and
\begin{equation}
T_H(x,y,Q^2) =  {16 \pi C_F \alpha_s(\mu)\over (1-x) (1-y)
Q^2}\left(1 + {\cal O}(\alpha_s)\right)
\end {equation}
is the leading-twist perturbatively-calculable subprocess amplitude
$\gamma^* q(x) \overline q(1-x) \to q(y) \overline q(1-y)$, obtained
by replacing the incident and final mesons by valence quarks collinear
up to the resolution scale ${\tilde Q}$.  The contributions from
non-valence Fock states and the correction from neglecting the
transverse momentum in the subprocess amplitude from the
non-perturbative region are higher twist, {\em i.e.}, power-law
suppressed. The transverse momenta in the perturbative domain lead to
the evolution of the distribution amplitude and to
next-to-leading-order (NLO) corrections in $\alpha_s$.  The
contribution from the endpoint regions of integration, $x \sim 1$ and
$y \sim 1,$ are power-law and Sudakov suppressed and thus can only
contribute corrections at higher order in $1/Q$ \cite{BrodskyLepage}.

The distribution amplitude $\phi(x,\tilde Q)$ is boost and gauge
invariant and evolves in $\ln \tilde Q$ through an evolution equation
\cite{BrodskyLepage}.  It can be computed from the integral over
transverse momenta of the renormalized hadron valence wavefunction in
the light-cone gauge at fixed light-cone time \cite{BrodskyLepage}:
\begin{equation}
\phi(x,\tilde Q) = \int d^2\vec{k_\perp}\thinspace
\theta \left({\tilde Q}^2 - {\vec{k_\perp}^2\over x(1-x)}\right)
\psi^{(\tilde Q)}(x,\vec{k_\perp}).
\label{quarkdistamp}
\end{equation}
The physical pion form factor must be independent of the separation
scale $\tilde Q.$ The natural variable in which to make this
separation is the light-cone energy, or equivalently the invariant
mass ${\cal M}^2 ={\vec{k_\perp}^2/ x(1-x)}$, of the off-shell
partonic system \cite{JIPang,BrodskyLepage}. Any residual dependence
on the choice of $\tilde Q$ for the distribution amplitude will be
compensated by a corresponding dependence of the NLO correction in
$T_H.$ However, the NLO prediction for the pion form factor depends
strongly on the form of the pion distribution amplitude as well as the
choice of renormalization scale $\mu$ and scheme.

It is straightforward to obtain the commensurate scale relation
between $F_\pi$ and $\alpha_V$ following the procedure outlined above.
The appropriate BLM scale for $F_\pi$ is determined from the explicit
calculations of the NLO corrections given by Dittes and Radyushkin
\cite{DittesRadyushkin} and Field {\em et al.} \cite{Field}.  These
may be written in the form $(A (\mu) n_f + B(\mu)) \alpha_s/\pi$,
where $A$ is independent of the separation scale $\tilde Q$.  The
$n_f$ dependence allows one to uniquely identify the dependence on
$\beta_0$, which is then absorbed into the running coupling by a shift
to the BLM scale $Q^*=e^{3A(\mu)}\mu$.  An important check of
self-consistency is that the resulting value for $Q^*$ is independent
of the choice of the starting scale $\mu$.

Combining this result with the BLM scale-fixed expression for
$\alpha_V$, and eliminating the intermediate coupling, we find
\begin{eqnarray}
F_\pi(Q^2)&=&\int^1_0 dx \phi_\pi(x)  \int^1_0  dy \phi_\pi(y)
{16 \pi C_F\alpha_V(Q_V)\over (1-x) (1-y) Q^2} \left(1 + C_V
{\alpha_V(Q_V)\over\pi}\right) \nonumber \\
\nonumber\\
&=& - 4\int^1_0 dx \phi_\pi(x)  \int^1_0  dy \phi_\pi(y)
V(Q_V^2) \left(1 + C_V {\alpha_V(Q_V)\over\pi}\right),
\label{pionformfactor}
\end{eqnarray}
where $C_V = -1.91$ is the same coefficient one would obtain in a
conformally invariant theory with $\beta=0$, and $Q_V^2 \equiv
(1-x)(1-y)Q^2$.  In this analysis we have assumed that the pion
distribution amplitude has the asymptotic form $\phi_\pi = {\sqrt 3}
f_\pi x (1-x)$, where the pion decay constant is $f_\pi \simeq 93$
MeV. In this simplified case the distribution amplitude does not
evolve, and there is no dependence on the separation scale $\tilde Q$.
This commensurate scale relation between $F_\pi(Q^2)$ and $\langle
\alpha_V(Q_V)\rangle$ represents a general connection between the form
factor of a bound-state system and the irreducible kernel that
describes the scattering of its constituents.

Alternatively, we can express the pion form factor in terms of other
effective charges such as the coupling $\alpha_R(\sqrt s)$ that
defines the QCD radiative corrections to the $e^+ e^- \to X$ cross
section: $R(s) \equiv 3\Sigma e_q^2 \left(1 + \alpha_R(\sqrt s)/
\pi\right).$ The CSR between $\alpha_V$ and $\alpha_R$ is
\begin{equation}
\alpha_{V}(Q_V) = \alpha_R(Q_R) \left(1 - {25\over 12}{
\alpha_R\over\pi} + \cdots\right),
\label{NLOalpv}
\end {equation}
where the ratio of commensurate scales to this order is $Q_R/Q_V =
e^{23/12-{2\zeta_3}} \simeq 0.614.$

If we expand the QCD coupling about a fixed point in NLO
\cite{LepageMackenzie}: $\alpha_s(Q_V) \simeq \alpha_s(Q_0) \left[1 -
\left(\beta_0\alpha_s(Q_0)/ 2\pi\right)\ln(Q_V/Q_0)\right]$, then the
integral over the effective charge in Eq. (\ref{pionformfactor}) can
be performed explicitly.  Thus, assuming the asymptotic distribution
amplitude, the pion form factor at NLO is
\begin{equation}
Q^2 F_\pi(Q^2) = 16 \pi f^2 _\pi  \alpha_V(Q^*) \left(1 - 1.91
{\alpha_V(Q^*)\over \pi}\right),
\label{pionformfactornlo}
\end{equation}
where $Q^* = e^{-3/2} Q$.  In this approximation $\ln{Q^*}^2 = \langle
\ln (1-x)(1-y)Q^2\rangle$, in agreement with the explicit calculation.
A striking feature of this result is that the physical scale
controlling the meson form factor in the $\alpha_V$ scheme is very
low: $e^{-3/2} Q \simeq 0.22 Q$, reflecting the characteristic
momentum transfer experienced by the spectator valence quark in
lepton-meson elastic scattering.

We may also determine the renormalization scale of $\alpha_V$ for more
general forms of the coupling by direct integration over $x$ and $y$
in Eq. (\ref{pionformfactor}), assuming a specific analytic form for
the coupling. Notice, however, that small corrections to the BLM scale
will be compensated by a corresponding change in the NLO coefficient.

Another exclusive amplitude of interest is the transition form factor
between a photon and a neutral hadron such as $F_{\gamma \pi}(Q^2)$,
which has now been measured up to $Q^2 < 8 $ GeV$^2$ in the tagged
two-photon collisions $e \gamma \to e' \pi^0$ by the CLEO and CELLO
collaborations.  In this case the amplitude has the factorized form
\begin{equation}
F_{\gamma M}(Q^2)= {4 \over \sqrt 3}\int^1_0 dx \phi_M(x,Q^2)
T^H_{\gamma \to M}(x,Q^2) ,
\label{transitionformfactor}
\end{equation}
where the hard scattering amplitude for $\gamma \gamma^* \to q \bar q$
is
\begin{equation}
T^H_{\gamma M}(x,Q^2) = {1\over (1-x) Q^2}\left(1 +
{\cal O}(\alpha_s)\right).
\label{transitionhardscattering}
\end{equation}
The leading QCD corrections have been computed by Braaten
\cite{Braaten}; however, the NLO corrections are necessary to fix the
BLM scale at LO.  Thus it is not yet possible to rigorously determine
the BLM scale for this quantity.  We shall here assume that this scale
is the same as that occurring in the prediction for $F_\pi$.  For the
asymptotic distribution amplitude we thus predict
\begin{equation}
Q^2 F_{\gamma \pi}(Q^2)= 2 f_\pi \left(1 - {5\over3}
{\alpha_V(Q^*)\over \pi}\right).
\label{qsquaretransition}
\end{equation}
As we shall see, given the phenomenological form of $\alpha_V$ we
employ (discussed below), this result is not terribly sensitive to the
precise value of the scale.

An important prediction resulting from the factorized form of these
results is that the normalization of the ratio
\begin{eqnarray}
R_\pi(Q^2) &\equiv & \frac{F_\pi (Q^2)}{4 \pi Q^2 |F_{\pi
\gamma}(Q^2)|^2}
\label{Rpidef}\\
&=& \alpha_{\overline{MS}}(e^{-14/6}Q)\left(1-0.56
{\alpha_{\overline{MS}} \over\pi} \right)
\label{Rpistart}\\
&=& \alpha_V(e^{-3/2}Q)\left(1+1.43 {\alpha_V\over\pi} \right)
\label{RpiV}\\
&=& \alpha_R(e^{5/12-{2\zeta_3}}Q) \left(1-
0.65 {\alpha_R\over\pi} \right)
\label{Rpiend}
\end{eqnarray}
is formally independent of the form of the pion distribution
amplitude. The $\alpha_{\overline{MS}}$ correction follows from
combined references \cite{DittesRadyushkin,Field,Braaten}.  The
next-to-leading correction given here assumes the asymptotic
distribution amplitude.

We emphasize that when we relate $R_\pi$ to $\alpha_V$ or $\alpha_R$
we relate observable to observable and thus there is no scheme
ambiguity.  The coefficients $-0.56$, $1.43$ and $-0.65$ in
Eqs. (\ref{Rpistart})--(\ref{Rpiend}) are identical to those one would
have in a theory with $\beta=0$, {\em i.e.}, conformally invariant
theory.

Contrary to the discussion by Chyla \cite{Chyla}, the optimized $Q^*$
is always {\em scheme} dependent.  For example, in the ${\overline
{MS}}$ scheme one finds $Q_{\overline {MS}}^2 = e^{-5/3} (1-x) (1-y)
Q^2$ for $F_\pi(Q^2)$ \cite{Field,BLM}, whereas in the $\alpha_V$
scheme the BLM scale is $Q_V^2=(1-x)(1-y)Q^2$.  The final results
connecting observables are of course scheme-independent.  The result
for $Q^2_V$ is expected since in the $\alpha_V$ scheme the scale of
the coupling is identified with the virtuality of the exchanged gluon
propagator, just as in the usual QED scheme, and here, to leading
twist, the virtuality of the gluon is $-(1-x)(1-y)Q^2$.  The resulting
relations between the form factors and the heavy quark coupling are
independent of the choice of intermediate renormalization scheme,
however; they thus have no scale or scheme ambiguities.

\section{The Behavior of the QCD Coupling at Low Momentum}

Effective charges such as $\alpha_V$ and $\alpha_R$ are defined from
physical observables and thus must be finite even at low momenta.  The
conventional solutions of the renormalization group equation for the
QCD coupling which are singular at $Q \simeq \Lambda_{\rm QCD}$ are
not accurate representations of the effective couplings at low
momentum transfer. It is clear that more parameters and information
are needed to specify the coupling in the non-perturbative domain.

A number of proposals have been suggested for the form of the QCD
coupling in the low-momentum regime.  For example, Parisi and
Petronzio \cite{PetronzioParisi} have argued that the coupling must
freeze at low momentum transfer in order that perturbative QCD loop
integrations be well defined.  
Similar ideas may be found in \cite{Greco}.
Mattingly and Stevenson
\cite{MattinglyStevenson} have incorporated such behavior into their
parameterizations of $\alpha_R$ at low scales.  Gribov \cite{Gribov}
has presented novel dynamical arguments related to the nature of
confinement for a fixed coupling at low scales.  Born
{\it et al.} \cite{Zerwas}
have noted the heavy quark potential must saturate to a Yukawa form
since the light-quark production processes will screen the linear
confining potential at large distances.  Cornwall \cite{Cornwall} and
others \cite{DonnachieLandshoff,GayDucatietal} have argued that the
gluon propagator will acquire an effective gluon mass $m_g$ from
non-perturbative dynamics, which again will regulate the form of the
effective couplings at low momentum. In this work we shall adopt the
simple parameterization
\begin{equation}
\alpha_V(Q) = {4 \pi \over {\beta_0 \ln \left({{Q^2 + 4m_g^2}\over
\Lambda^2_V}\right)}} ,
\label{frozencoupling}
\end{equation}
which effectively freezes the $\alpha_V$ effective charge to a finite
value for $Q^2 \leq 4m_g^2.$

\begin{figure}
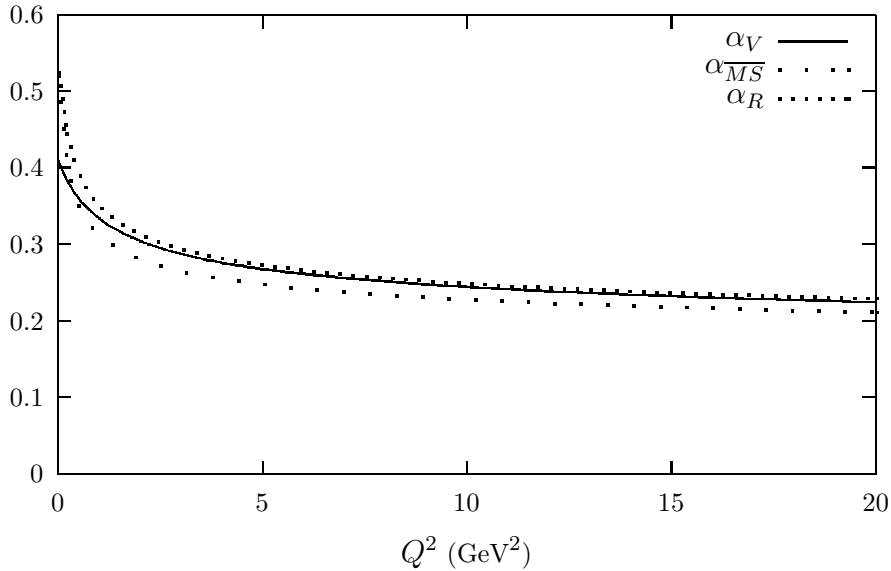

\begin{center}
\input fig1.tex
\end{center}
\caption{The coupling function $\alpha_V(Q^2)$ as given in
Eq. (\ref{frozencoupling}).  Also shown are the corresponding
predictions for $\alpha_{\overline{MS}}$ and $\alpha_R$ following from
the NLO commensurate scale relations [Eqs. (\ref{alpmsbar}) and
(\ref{NLOalpv})].}
\label{couplings}
\end{figure}

We can use the non-relativistic heavy quark lattice results
\cite{Davies,Sloan} to fix the parameters.  A fit to the lattice data
of the above parameterization gives $\Lambda_V = 0.16$ GeV if we use
the well-known momentum-dependent $n_f$ \cite{ShirkovMikhailov}.
Furthermore, the value $m^2_g=0.19$ GeV$^2$ gives consistency with the
frozen value of $\alpha_R$ advocated by Mattingly and Stevenson
\cite{MattinglyStevenson}.  Their parameterization implies the
approximate constraint $\alpha_R(Q)/\pi \simeq 0.27$ for $Q= \sqrt s <
0.3$ GeV, which leads to $\alpha_V(0.5~{\rm GeV}) \simeq 0.37$ using
the NLO commensurate scale relation between $\alpha_V$ and $\alpha_R$.
The resulting form for $\alpha_V$ is shown in Fig. \ref{couplings}.
The corresponding predictions for $\alpha_R$ and $\alpha_{\overline
{MS}}$ using the CSRs at NLO are also shown.  Note that for low $Q^2$
the couplings, although frozen, are large.  Thus the NLO and
higher-order terms in the CSRs are large, and inverting them
perturbatively to NLO does not give accurate results at low scales.
In addition, higher-twist contributions to $\alpha_V$ and $\alpha_R$,
which are not reflected in the CSR relating them, may be expected to
be important for low $Q^2$ \cite{Braun}.

It is clear that exclusive processes such as the pion and photon to
pion transition form factors can provide a valuable window for
determining the magnitude and the shape of the effective charges at
quite low momentum transfers.  In particular, we can check consistency
with the $\alpha_V$ prediction from lattice gauge theory.  A
complimentary method for determining $\alpha_V$ at low momentum is to
use the angular anisotropy of $e^+ e^- \to Q \overline Q$ at the heavy
quark thresholds \cite{BrodskyKuhnHoangTuebner}. It should be
emphasized that the parameterization (\ref{frozencoupling}) is just an
approximate form. The actual behavior of $\alpha_V(Q^2)$ at low $Q^2$
is one of the key uncertainties in QCD phenomenology.  In this paper
we shall use exclusive observables to deduce information on this
quantity.
\vfill\pagebreak

\section{Applications}

As we have emphasized, exclusive processes are sensitive to the
magnitude and shape of the QCD couplings at quite low momentum
transfer: $Q_V^{*2} \simeq e^{-3} Q^2 \simeq Q^2/20$ and $Q_R^{*2}
\simeq Q^2/50$ \cite{LewellynIsgur}.  The fact that the data for
exclusive processes such as form factors, two photon processes such as
$\gamma \gamma \to \pi^+ \pi^-,$ and photoproduction at fixed
$\theta_{c.m.}$ are consistent with the nominal scaling of the
leading-twist QCD predictions (dimensional counting) at momentum
transfers $Q$ up to the order of a few GeV can be immediately
understood if the effective charges $\alpha_V$ and $\alpha_R$ are
slowly varying at low momentum.  The scaling of the exclusive
amplitude then follows that of the subprocess amplitude $T_H$ with
effectively fixed coupling. Note also that the Sudakov effect of the
endpoint region is the exponential of a double log series if the
coupling is frozen, and thus is strong.

\begin{figure}
\begin{center}
\setlength{\unitlength}{0.240900pt}
\ifx\plotpoint\undefined\newsavebox{\plotpoint}\fi
\sbox{\plotpoint}{\rule[-0.200pt]{0.400pt}{0.400pt}}%
\begin{picture}(1500,900)(0,0)
\font\gnuplot=cmr10 at 10pt
\gnuplot
\sbox{\plotpoint}{\rule[-0.200pt]{0.400pt}{0.400pt}}%
\put(219.0,134.0){\rule[-0.200pt]{4.818pt}{0.400pt}}
\put(197,134){\makebox(0,0)[r]{0}}
\put(1416.0,134.0){\rule[-0.200pt]{4.818pt}{0.400pt}}
\put(219.0,278.0){\rule[-0.200pt]{4.818pt}{0.400pt}}
\put(197,278){\makebox(0,0)[r]{0.05}}
\put(1416.0,278.0){\rule[-0.200pt]{4.818pt}{0.400pt}}
\put(219.0,422.0){\rule[-0.200pt]{4.818pt}{0.400pt}}
\put(197,422){\makebox(0,0)[r]{0.1}}
\put(1416.0,422.0){\rule[-0.200pt]{4.818pt}{0.400pt}}
\put(219.0,567.0){\rule[-0.200pt]{4.818pt}{0.400pt}}
\put(197,567){\makebox(0,0)[r]{0.15}}
\put(1416.0,567.0){\rule[-0.200pt]{4.818pt}{0.400pt}}
\put(219.0,711.0){\rule[-0.200pt]{4.818pt}{0.400pt}}
\put(197,711){\makebox(0,0)[r]{0.2}}
\put(1416.0,711.0){\rule[-0.200pt]{4.818pt}{0.400pt}}
\put(219.0,855.0){\rule[-0.200pt]{4.818pt}{0.400pt}}
\put(197,855){\makebox(0,0)[r]{0.25}}
\put(1416.0,855.0){\rule[-0.200pt]{4.818pt}{0.400pt}}
\put(219.0,134.0){\rule[-0.200pt]{0.400pt}{4.818pt}}
\put(219,89){\makebox(0,0){0}}
\put(219.0,835.0){\rule[-0.200pt]{0.400pt}{4.818pt}}
\put(462.0,134.0){\rule[-0.200pt]{0.400pt}{4.818pt}}
\put(462,89){\makebox(0,0){2}}
\put(462.0,835.0){\rule[-0.200pt]{0.400pt}{4.818pt}}
\put(706.0,134.0){\rule[-0.200pt]{0.400pt}{4.818pt}}
\put(706,89){\makebox(0,0){4}}
\put(706.0,835.0){\rule[-0.200pt]{0.400pt}{4.818pt}}
\put(949.0,134.0){\rule[-0.200pt]{0.400pt}{4.818pt}}
\put(949,89){\makebox(0,0){6}}
\put(949.0,835.0){\rule[-0.200pt]{0.400pt}{4.818pt}}
\put(1193.0,134.0){\rule[-0.200pt]{0.400pt}{4.818pt}}
\put(1193,89){\makebox(0,0){8}}
\put(1193.0,835.0){\rule[-0.200pt]{0.400pt}{4.818pt}}
\put(1436.0,134.0){\rule[-0.200pt]{0.400pt}{4.818pt}}
\put(1436,89){\makebox(0,0){10}}
\put(1436.0,835.0){\rule[-0.200pt]{0.400pt}{4.818pt}}
\put(219.0,134.0){\rule[-0.200pt]{293.175pt}{0.400pt}}
\put(1436.0,134.0){\rule[-0.200pt]{0.400pt}{173.689pt}}
\put(219.0,855.0){\rule[-0.200pt]{293.175pt}{0.400pt}}
\put(-22,494){\makebox(0,0){\shortstack{$Q^2 F_{\gamma\pi} (Q^2)$\\ \\ (GeV)}}}
\put(827,10){\makebox(0,0){$Q^2$ (GeV$^2$)}}
\put(219.0,134.0){\rule[-0.200pt]{0.400pt}{173.689pt}}
\put(419,483){\circle*{12}}
\put(450,471){\circle*{12}}
\put(475,532){\circle*{12}}
\put(499,500){\circle*{12}}
\put(523,523){\circle*{12}}
\put(548,569){\circle*{12}}
\put(577,529){\circle*{12}}
\put(619,552){\circle*{12}}
\put(674,515){\circle*{12}}
\put(735,520){\circle*{12}}
\put(796,578){\circle*{12}}
\put(857,552){\circle*{12}}
\put(918,581){\circle*{12}}
\put(1006,561){\circle*{12}}
\put(1180,616){\circle*{12}}
\put(419.0,451.0){\rule[-0.200pt]{0.400pt}{15.418pt}}
\put(409.0,451.0){\rule[-0.200pt]{4.818pt}{0.400pt}}
\put(409.0,515.0){\rule[-0.200pt]{4.818pt}{0.400pt}}
\put(450.0,443.0){\rule[-0.200pt]{0.400pt}{13.731pt}}
\put(440.0,443.0){\rule[-0.200pt]{4.818pt}{0.400pt}}
\put(440.0,500.0){\rule[-0.200pt]{4.818pt}{0.400pt}}
\put(475.0,500.0){\rule[-0.200pt]{0.400pt}{15.418pt}}
\put(465.0,500.0){\rule[-0.200pt]{4.818pt}{0.400pt}}
\put(465.0,564.0){\rule[-0.200pt]{4.818pt}{0.400pt}}
\put(499.0,466.0){\rule[-0.200pt]{0.400pt}{16.622pt}}
\put(489.0,466.0){\rule[-0.200pt]{4.818pt}{0.400pt}}
\put(489.0,535.0){\rule[-0.200pt]{4.818pt}{0.400pt}}
\put(523.0,486.0){\rule[-0.200pt]{0.400pt}{18.067pt}}
\put(513.0,486.0){\rule[-0.200pt]{4.818pt}{0.400pt}}
\put(513.0,561.0){\rule[-0.200pt]{4.818pt}{0.400pt}}
\put(548.0,526.0){\rule[-0.200pt]{0.400pt}{20.958pt}}
\put(538.0,526.0){\rule[-0.200pt]{4.818pt}{0.400pt}}
\put(538.0,613.0){\rule[-0.200pt]{4.818pt}{0.400pt}}
\put(577.0,486.0){\rule[-0.200pt]{0.400pt}{20.717pt}}
\put(567.0,486.0){\rule[-0.200pt]{4.818pt}{0.400pt}}
\put(567.0,572.0){\rule[-0.200pt]{4.818pt}{0.400pt}}
\put(619.0,506.0){\rule[-0.200pt]{0.400pt}{22.163pt}}
\put(609.0,506.0){\rule[-0.200pt]{4.818pt}{0.400pt}}
\put(609.0,598.0){\rule[-0.200pt]{4.818pt}{0.400pt}}
\put(674.0,466.0){\rule[-0.200pt]{0.400pt}{23.608pt}}
\put(664.0,466.0){\rule[-0.200pt]{4.818pt}{0.400pt}}
\put(664.0,564.0){\rule[-0.200pt]{4.818pt}{0.400pt}}
\put(735.0,469.0){\rule[-0.200pt]{0.400pt}{24.813pt}}
\put(725.0,469.0){\rule[-0.200pt]{4.818pt}{0.400pt}}
\put(725.0,572.0){\rule[-0.200pt]{4.818pt}{0.400pt}}
\put(796.0,518.0){\rule[-0.200pt]{0.400pt}{29.149pt}}
\put(786.0,518.0){\rule[-0.200pt]{4.818pt}{0.400pt}}
\put(786.0,639.0){\rule[-0.200pt]{4.818pt}{0.400pt}}
\put(857.0,489.0){\rule[-0.200pt]{0.400pt}{30.594pt}}
\put(847.0,489.0){\rule[-0.200pt]{4.818pt}{0.400pt}}
\put(847.0,616.0){\rule[-0.200pt]{4.818pt}{0.400pt}}
\put(918.0,506.0){\rule[-0.200pt]{0.400pt}{36.135pt}}
\put(908.0,506.0){\rule[-0.200pt]{4.818pt}{0.400pt}}
\put(908.0,656.0){\rule[-0.200pt]{4.818pt}{0.400pt}}
\put(1006.0,492.0){\rule[-0.200pt]{0.400pt}{33.244pt}}
\put(996.0,492.0){\rule[-0.200pt]{4.818pt}{0.400pt}}
\put(996.0,630.0){\rule[-0.200pt]{4.818pt}{0.400pt}}
\put(1180.0,532.0){\rule[-0.200pt]{0.400pt}{40.230pt}}
\put(1170.0,532.0){\rule[-0.200pt]{4.818pt}{0.400pt}}
\put(1170.0,699.0){\rule[-0.200pt]{4.818pt}{0.400pt}}
\sbox{\plotpoint}{\rule[-0.500pt]{1.000pt}{1.000pt}}%
\put(341,670){\usebox{\plotpoint}}
\multiput(341,670)(20.756,0.000){52}{\usebox{\plotpoint}}
\put(1418,670){\usebox{\plotpoint}}
\sbox{\plotpoint}{\rule[-0.200pt]{0.400pt}{0.400pt}}%
\put(341,561){\usebox{\plotpoint}}
\put(430,560.67){\rule{5.300pt}{0.400pt}}
\multiput(430.00,560.17)(11.000,1.000){2}{\rule{2.650pt}{0.400pt}}
\put(341.0,561.0){\rule[-0.200pt]{21.440pt}{0.400pt}}
\put(587,561.67){\rule{5.300pt}{0.400pt}}
\multiput(587.00,561.17)(11.000,1.000){2}{\rule{2.650pt}{0.400pt}}
\put(452.0,562.0){\rule[-0.200pt]{32.521pt}{0.400pt}}
\put(698,562.67){\rule{5.541pt}{0.400pt}}
\multiput(698.00,562.17)(11.500,1.000){2}{\rule{2.770pt}{0.400pt}}
\put(609.0,563.0){\rule[-0.200pt]{21.440pt}{0.400pt}}
\put(810,563.67){\rule{5.300pt}{0.400pt}}
\multiput(810.00,563.17)(11.000,1.000){2}{\rule{2.650pt}{0.400pt}}
\put(721.0,564.0){\rule[-0.200pt]{21.440pt}{0.400pt}}
\put(944,564.67){\rule{5.541pt}{0.400pt}}
\multiput(944.00,564.17)(11.500,1.000){2}{\rule{2.770pt}{0.400pt}}
\put(832.0,565.0){\rule[-0.200pt]{26.981pt}{0.400pt}}
\put(1056,565.67){\rule{5.300pt}{0.400pt}}
\multiput(1056.00,565.17)(11.000,1.000){2}{\rule{2.650pt}{0.400pt}}
\put(967.0,566.0){\rule[-0.200pt]{21.440pt}{0.400pt}}
\put(1190,566.67){\rule{5.300pt}{0.400pt}}
\multiput(1190.00,566.17)(11.000,1.000){2}{\rule{2.650pt}{0.400pt}}
\put(1078.0,567.0){\rule[-0.200pt]{26.981pt}{0.400pt}}
\put(1324,567.67){\rule{5.541pt}{0.400pt}}
\multiput(1324.00,567.17)(11.500,1.000){2}{\rule{2.770pt}{0.400pt}}
\put(1212.0,568.0){\rule[-0.200pt]{26.981pt}{0.400pt}}
\put(1347.0,569.0){\rule[-0.200pt]{21.440pt}{0.400pt}}
\end{picture}
\end{center}
\caption{The $\gamma\rightarrow\pi^0$ transition form factor.  The
solid line is the full prediction including the QCD correction
[Eq. (\ref{fgampi})]; the dotted line is the LO prediction
$Q^2F_{\gamma\pi}(Q^2) = 2f_\pi$.}
\label{fgammapi}
\end{figure}
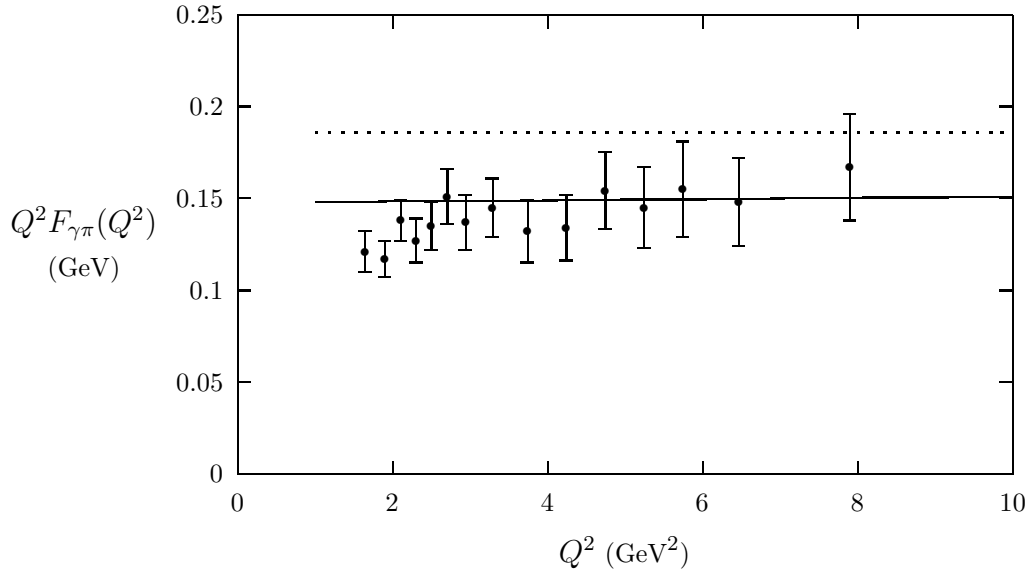

In Fig. \ref{fgammapi}, we compare the recent CLEO
data \cite{Cleo} for the photon to pion transition form factor with
the prediction
\begin{equation}
Q^2 F_{\gamma\pi}(Q^2)= 2 f_\pi \left( 1
- {5\over3} {\alpha_V(e^{-3/2} Q)\over \pi}\right).
\label{fgampi}
\end{equation}
The flat scaling of the $Q^2 F_{\gamma \pi}(Q^2)$ data from $Q^2 = 2$
to $Q^2 = 8$ GeV$^2$ provides an important confirmation of the
applicability of leading twist QCD to this process. The magnitude of
$Q^2 F_{\gamma \pi}(Q^2)$ is remarkably consistent with the predicted
form assuming the asymptotic distribution amplitude and including the
LO QCD radiative correction with $\alpha_V(e^{-3/2} Q)/\pi \simeq
0.12$.  Radyushkin \cite {Radyushkin}, Ong \cite{Ong} and Kroll \cite
{Kroll} have also noted that the scaling and normalization of the
photon-to-pion transition form factor tends to favor the asymptotic
form for the pion distribution amplitude and rules out broader
distributions such as the two-humped form suggested by QCD sum rules
\cite{CZ}.  One cannot obtain a unique solution for the
non-perturbative wavefunction from the $F_{\pi\gamma}$ data alone.
However, we have the constraint that
\begin{equation}
{1\over 3}\Biggl\langle {1\over 1-x}\Biggl\rangle \left[ 1-{5\over
3}{\alpha_V(Q^*)\over\pi} \right]\simeq 0.8
\end{equation}
(assuming the renormalization scale we have chosen in
Eq. (\ref{qsquaretransition}) is approximately correct).  Thus one
could allow for some broadening of the distribution amplitude with a
corresponding increase in the value of $\alpha_V$ at low scales.

\begin{figure}
\begin{center}
\setlength{\unitlength}{0.240900pt}
\ifx\plotpoint\undefined\newsavebox{\plotpoint}\fi
\sbox{\plotpoint}{\rule[-0.200pt]{0.400pt}{0.400pt}}%
\begin{picture}(1500,900)(0,0)
\font\gnuplot=cmr10 at 10pt
\gnuplot
\sbox{\plotpoint}{\rule[-0.200pt]{0.400pt}{0.400pt}}%
\put(197.0,134.0){\rule[-0.200pt]{4.818pt}{0.400pt}}
\put(175,134){\makebox(0,0)[r]{0}}
\put(1416.0,134.0){\rule[-0.200pt]{4.818pt}{0.400pt}}
\put(197.0,254.0){\rule[-0.200pt]{4.818pt}{0.400pt}}
\put(175,254){\makebox(0,0)[r]{0.1}}
\put(1416.0,254.0){\rule[-0.200pt]{4.818pt}{0.400pt}}
\put(197.0,374.0){\rule[-0.200pt]{4.818pt}{0.400pt}}
\put(175,374){\makebox(0,0)[r]{0.2}}
\put(1416.0,374.0){\rule[-0.200pt]{4.818pt}{0.400pt}}
\put(197.0,495.0){\rule[-0.200pt]{4.818pt}{0.400pt}}
\put(175,495){\makebox(0,0)[r]{0.3}}
\put(1416.0,495.0){\rule[-0.200pt]{4.818pt}{0.400pt}}
\put(197.0,615.0){\rule[-0.200pt]{4.818pt}{0.400pt}}
\put(175,615){\makebox(0,0)[r]{0.4}}
\put(1416.0,615.0){\rule[-0.200pt]{4.818pt}{0.400pt}}
\put(197.0,735.0){\rule[-0.200pt]{4.818pt}{0.400pt}}
\put(175,735){\makebox(0,0)[r]{0.5}}
\put(1416.0,735.0){\rule[-0.200pt]{4.818pt}{0.400pt}}
\put(197.0,855.0){\rule[-0.200pt]{4.818pt}{0.400pt}}
\put(175,855){\makebox(0,0)[r]{0.6}}
\put(1416.0,855.0){\rule[-0.200pt]{4.818pt}{0.400pt}}
\put(197.0,134.0){\rule[-0.200pt]{0.400pt}{4.818pt}}
\put(197,89){\makebox(0,0){0}}
\put(197.0,835.0){\rule[-0.200pt]{0.400pt}{4.818pt}}
\put(445.0,134.0){\rule[-0.200pt]{0.400pt}{4.818pt}}
\put(445,89){\makebox(0,0){2}}
\put(445.0,835.0){\rule[-0.200pt]{0.400pt}{4.818pt}}
\put(693.0,134.0){\rule[-0.200pt]{0.400pt}{4.818pt}}
\put(693,89){\makebox(0,0){4}}
\put(693.0,835.0){\rule[-0.200pt]{0.400pt}{4.818pt}}
\put(940.0,134.0){\rule[-0.200pt]{0.400pt}{4.818pt}}
\put(940,89){\makebox(0,0){6}}
\put(940.0,835.0){\rule[-0.200pt]{0.400pt}{4.818pt}}
\put(1188.0,134.0){\rule[-0.200pt]{0.400pt}{4.818pt}}
\put(1188,89){\makebox(0,0){8}}
\put(1188.0,835.0){\rule[-0.200pt]{0.400pt}{4.818pt}}
\put(1436.0,134.0){\rule[-0.200pt]{0.400pt}{4.818pt}}
\put(1436,89){\makebox(0,0){10}}
\put(1436.0,835.0){\rule[-0.200pt]{0.400pt}{4.818pt}}
\put(197.0,134.0){\rule[-0.200pt]{298.475pt}{0.400pt}}
\put(1436.0,134.0){\rule[-0.200pt]{0.400pt}{173.689pt}}
\put(197.0,855.0){\rule[-0.200pt]{298.475pt}{0.400pt}}
\put(-22,494){\makebox(0,0){\shortstack{$Q^2 F_{\pi} (Q^2)$\\ \\ (GeV$^2$)}}}
\put(816,10){\makebox(0,0){$Q^2$ (GeV$^2$)}}
\put(197.0,134.0){\rule[-0.200pt]{0.400pt}{173.689pt}}
\put(219,318){\circle*{12}}
\put(233,355){\circle*{12}}
\put(247,408){\circle*{12}}
\put(295,499){\circle*{12}}
\put(344,474){\circle*{12}}
\put(274,466){\circle*{12}}
\put(330,531){\circle*{12}}
\put(346,522){\circle*{12}}
\put(359,515){\circle*{12}}
\put(346,512){\circle*{12}}
\put(446,506){\circle*{12}}
\put(348,559){\circle*{12}}
\put(346,558){\circle*{12}}
\put(409,623){\circle*{12}}
\put(606,538){\circle*{12}}
\put(444,562){\circle*{12}}
\put(343,497){\circle*{12}}
\put(437,584){\circle*{12}}
\put(610,478){\circle*{12}}
\put(978,581){\circle*{12}}
\put(219.0,308.0){\rule[-0.200pt]{0.400pt}{4.577pt}}
\put(209.0,308.0){\rule[-0.200pt]{4.818pt}{0.400pt}}
\put(209.0,327.0){\rule[-0.200pt]{4.818pt}{0.400pt}}
\put(233.0,345.0){\rule[-0.200pt]{0.400pt}{4.818pt}}
\put(223.0,345.0){\rule[-0.200pt]{4.818pt}{0.400pt}}
\put(223.0,365.0){\rule[-0.200pt]{4.818pt}{0.400pt}}
\put(247.0,400.0){\rule[-0.200pt]{0.400pt}{3.854pt}}
\put(237.0,400.0){\rule[-0.200pt]{4.818pt}{0.400pt}}
\put(237.0,416.0){\rule[-0.200pt]{4.818pt}{0.400pt}}
\put(295.0,485.0){\rule[-0.200pt]{0.400pt}{6.504pt}}
\put(285.0,485.0){\rule[-0.200pt]{4.818pt}{0.400pt}}
\put(285.0,512.0){\rule[-0.200pt]{4.818pt}{0.400pt}}
\put(344.0,450.0){\rule[-0.200pt]{0.400pt}{11.804pt}}
\put(334.0,450.0){\rule[-0.200pt]{4.818pt}{0.400pt}}
\put(334.0,499.0){\rule[-0.200pt]{4.818pt}{0.400pt}}
\put(274.0,454.0){\rule[-0.200pt]{0.400pt}{5.541pt}}
\put(264.0,454.0){\rule[-0.200pt]{4.818pt}{0.400pt}}
\put(264.0,477.0){\rule[-0.200pt]{4.818pt}{0.400pt}}
\put(330.0,507.0){\rule[-0.200pt]{0.400pt}{11.804pt}}
\put(320.0,507.0){\rule[-0.200pt]{4.818pt}{0.400pt}}
\put(320.0,556.0){\rule[-0.200pt]{4.818pt}{0.400pt}}
\put(346.0,505.0){\rule[-0.200pt]{0.400pt}{8.191pt}}
\put(336.0,505.0){\rule[-0.200pt]{4.818pt}{0.400pt}}
\put(336.0,539.0){\rule[-0.200pt]{4.818pt}{0.400pt}}
\put(359.0,491.0){\rule[-0.200pt]{0.400pt}{11.563pt}}
\put(349.0,491.0){\rule[-0.200pt]{4.818pt}{0.400pt}}
\put(349.0,539.0){\rule[-0.200pt]{4.818pt}{0.400pt}}
\put(346.0,492.0){\rule[-0.200pt]{0.400pt}{9.636pt}}
\put(336.0,492.0){\rule[-0.200pt]{4.818pt}{0.400pt}}
\put(336.0,532.0){\rule[-0.200pt]{4.818pt}{0.400pt}}
\put(446.0,472.0){\rule[-0.200pt]{0.400pt}{16.381pt}}
\put(436.0,472.0){\rule[-0.200pt]{4.818pt}{0.400pt}}
\put(436.0,540.0){\rule[-0.200pt]{4.818pt}{0.400pt}}
\put(348.0,515.0){\rule[-0.200pt]{0.400pt}{21.199pt}}
\put(338.0,515.0){\rule[-0.200pt]{4.818pt}{0.400pt}}
\put(338.0,603.0){\rule[-0.200pt]{4.818pt}{0.400pt}}
\put(346.0,531.0){\rule[-0.200pt]{0.400pt}{13.009pt}}
\put(336.0,531.0){\rule[-0.200pt]{4.818pt}{0.400pt}}
\put(336.0,585.0){\rule[-0.200pt]{4.818pt}{0.400pt}}
\put(409.0,582.0){\rule[-0.200pt]{0.400pt}{19.754pt}}
\put(399.0,582.0){\rule[-0.200pt]{4.818pt}{0.400pt}}
\put(399.0,664.0){\rule[-0.200pt]{4.818pt}{0.400pt}}
\put(606.0,447.0){\rule[-0.200pt]{0.400pt}{44.085pt}}
\put(596.0,447.0){\rule[-0.200pt]{4.818pt}{0.400pt}}
\put(596.0,630.0){\rule[-0.200pt]{4.818pt}{0.400pt}}
\put(444.0,512.0){\rule[-0.200pt]{0.400pt}{24.090pt}}
\put(434.0,512.0){\rule[-0.200pt]{4.818pt}{0.400pt}}
\put(434.0,612.0){\rule[-0.200pt]{4.818pt}{0.400pt}}
\put(343.0,460.0){\rule[-0.200pt]{0.400pt}{17.827pt}}
\put(333.0,460.0){\rule[-0.200pt]{4.818pt}{0.400pt}}
\put(333.0,534.0){\rule[-0.200pt]{4.818pt}{0.400pt}}
\put(437.0,526.0){\rule[-0.200pt]{0.400pt}{27.944pt}}
\put(427.0,526.0){\rule[-0.200pt]{4.818pt}{0.400pt}}
\put(427.0,642.0){\rule[-0.200pt]{4.818pt}{0.400pt}}
\put(610.0,346.0){\rule[-0.200pt]{0.400pt}{63.598pt}}
\put(600.0,346.0){\rule[-0.200pt]{4.818pt}{0.400pt}}
\put(600.0,610.0){\rule[-0.200pt]{4.818pt}{0.400pt}}
\put(978.0,354.0){\rule[-0.200pt]{0.400pt}{109.369pt}}
\put(968.0,354.0){\rule[-0.200pt]{4.818pt}{0.400pt}}
\put(968.0,808.0){\rule[-0.200pt]{4.818pt}{0.400pt}}
\put(321,288){\usebox{\plotpoint}}
\put(412,286.67){\rule{5.541pt}{0.400pt}}
\multiput(412.00,287.17)(11.500,-1.000){2}{\rule{2.770pt}{0.400pt}}
\put(321.0,288.0){\rule[-0.200pt]{21.922pt}{0.400pt}}
\put(548,285.67){\rule{5.541pt}{0.400pt}}
\multiput(548.00,286.17)(11.500,-1.000){2}{\rule{2.770pt}{0.400pt}}
\put(435.0,287.0){\rule[-0.200pt]{27.222pt}{0.400pt}}
\put(685,284.67){\rule{5.541pt}{0.400pt}}
\multiput(685.00,285.17)(11.500,-1.000){2}{\rule{2.770pt}{0.400pt}}
\put(571.0,286.0){\rule[-0.200pt]{27.463pt}{0.400pt}}
\put(799,283.67){\rule{5.541pt}{0.400pt}}
\multiput(799.00,284.17)(11.500,-1.000){2}{\rule{2.770pt}{0.400pt}}
\put(708.0,285.0){\rule[-0.200pt]{21.922pt}{0.400pt}}
\put(913,282.67){\rule{5.300pt}{0.400pt}}
\multiput(913.00,283.17)(11.000,-1.000){2}{\rule{2.650pt}{0.400pt}}
\put(822.0,284.0){\rule[-0.200pt]{21.922pt}{0.400pt}}
\put(1049,281.67){\rule{5.541pt}{0.400pt}}
\multiput(1049.00,282.17)(11.500,-1.000){2}{\rule{2.770pt}{0.400pt}}
\put(935.0,283.0){\rule[-0.200pt]{27.463pt}{0.400pt}}
\put(1163,280.67){\rule{5.541pt}{0.400pt}}
\multiput(1163.00,281.17)(11.500,-1.000){2}{\rule{2.770pt}{0.400pt}}
\put(1072.0,282.0){\rule[-0.200pt]{21.922pt}{0.400pt}}
\put(1299,279.67){\rule{5.541pt}{0.400pt}}
\multiput(1299.00,280.17)(11.500,-1.000){2}{\rule{2.770pt}{0.400pt}}
\put(1186.0,281.0){\rule[-0.200pt]{27.222pt}{0.400pt}}
\put(1322.0,280.0){\rule[-0.200pt]{27.463pt}{0.400pt}}
\end{picture}
\end{center}
\caption{The space-like pion form factor.}
\label{fpi}
\end{figure}
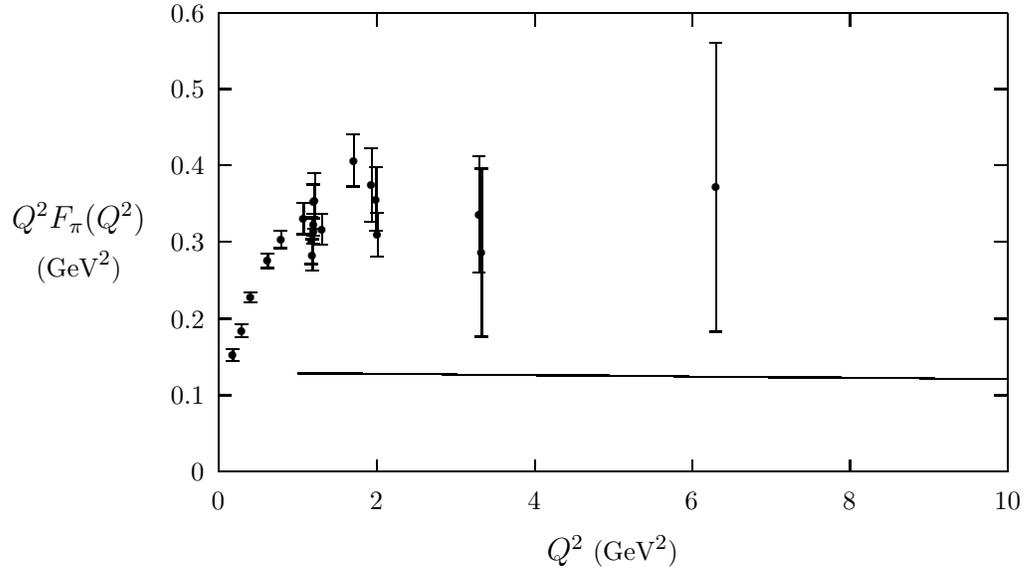

\begin{figure}
\begin{center}
\setlength{\unitlength}{0.240900pt}
\ifx\plotpoint\undefined\newsavebox{\plotpoint}\fi
\sbox{\plotpoint}{\rule[-0.200pt]{0.400pt}{0.400pt}}%
\begin{picture}(1500,900)(0,0)
\font\gnuplot=cmr10 at 10pt
\gnuplot
\sbox{\plotpoint}{\rule[-0.200pt]{0.400pt}{0.400pt}}%
\put(197.0,134.0){\rule[-0.200pt]{4.818pt}{0.400pt}}
\put(175,134){\makebox(0,0)[r]{0}}
\put(1416.0,134.0){\rule[-0.200pt]{4.818pt}{0.400pt}}
\put(197.0,314.0){\rule[-0.200pt]{4.818pt}{0.400pt}}
\put(175,314){\makebox(0,0)[r]{0.5}}
\put(1416.0,314.0){\rule[-0.200pt]{4.818pt}{0.400pt}}
\put(197.0,495.0){\rule[-0.200pt]{4.818pt}{0.400pt}}
\put(175,495){\makebox(0,0)[r]{1}}
\put(1416.0,495.0){\rule[-0.200pt]{4.818pt}{0.400pt}}
\put(197.0,675.0){\rule[-0.200pt]{4.818pt}{0.400pt}}
\put(175,675){\makebox(0,0)[r]{1.5}}
\put(1416.0,675.0){\rule[-0.200pt]{4.818pt}{0.400pt}}
\put(197.0,855.0){\rule[-0.200pt]{4.818pt}{0.400pt}}
\put(175,855){\makebox(0,0)[r]{2}}
\put(1416.0,855.0){\rule[-0.200pt]{4.818pt}{0.400pt}}
\put(197.0,134.0){\rule[-0.200pt]{0.400pt}{4.818pt}}
\put(197,89){\makebox(0,0){0}}
\put(197.0,835.0){\rule[-0.200pt]{0.400pt}{4.818pt}}
\put(445.0,134.0){\rule[-0.200pt]{0.400pt}{4.818pt}}
\put(445,89){\makebox(0,0){2}}
\put(445.0,835.0){\rule[-0.200pt]{0.400pt}{4.818pt}}
\put(693.0,134.0){\rule[-0.200pt]{0.400pt}{4.818pt}}
\put(693,89){\makebox(0,0){4}}
\put(693.0,835.0){\rule[-0.200pt]{0.400pt}{4.818pt}}
\put(940.0,134.0){\rule[-0.200pt]{0.400pt}{4.818pt}}
\put(940,89){\makebox(0,0){6}}
\put(940.0,835.0){\rule[-0.200pt]{0.400pt}{4.818pt}}
\put(1188.0,134.0){\rule[-0.200pt]{0.400pt}{4.818pt}}
\put(1188,89){\makebox(0,0){8}}
\put(1188.0,835.0){\rule[-0.200pt]{0.400pt}{4.818pt}}
\put(1436.0,134.0){\rule[-0.200pt]{0.400pt}{4.818pt}}
\put(1436,89){\makebox(0,0){10}}
\put(1436.0,835.0){\rule[-0.200pt]{0.400pt}{4.818pt}}
\put(197.0,134.0){\rule[-0.200pt]{298.475pt}{0.400pt}}
\put(1436.0,134.0){\rule[-0.200pt]{0.400pt}{173.689pt}}
\put(197.0,855.0){\rule[-0.200pt]{298.475pt}{0.400pt}}
\put(22,494){\makebox(0,0){$R_{\pi} (Q^2)$}}
\put(816,10){\makebox(0,0){$Q^2$ (GeV$^2$)}}
\put(197.0,134.0){\rule[-0.200pt]{0.400pt}{173.689pt}}
\put(458,497){\circle*{12}}
\put(602,502){\circle*{12}}
\put(990,615){\circle*{12}}
\put(458.0,263.0){\rule[-0.200pt]{0.400pt}{142.613pt}}
\put(448.0,263.0){\rule[-0.200pt]{4.818pt}{0.400pt}}
\put(448.0,855.0){\rule[-0.200pt]{4.818pt}{0.400pt}}
\put(602.0,313.0){\rule[-0.200pt]{0.400pt}{130.568pt}}
\put(592.0,313.0){\rule[-0.200pt]{4.818pt}{0.400pt}}
\put(592.0,855.0){\rule[-0.200pt]{4.818pt}{0.400pt}}
\put(990.0,442.0){\rule[-0.200pt]{0.400pt}{99.492pt}}
\put(980.0,442.0){\rule[-0.200pt]{4.818pt}{0.400pt}}
\put(980.0,855.0){\rule[-0.200pt]{4.818pt}{0.400pt}}
\put(321,297){\usebox{\plotpoint}}
\put(366,295.67){\rule{5.541pt}{0.400pt}}
\multiput(366.00,296.17)(11.500,-1.000){2}{\rule{2.770pt}{0.400pt}}
\put(321.0,297.0){\rule[-0.200pt]{10.840pt}{0.400pt}}
\put(480,294.67){\rule{5.541pt}{0.400pt}}
\multiput(480.00,295.17)(11.500,-1.000){2}{\rule{2.770pt}{0.400pt}}
\put(389.0,296.0){\rule[-0.200pt]{21.922pt}{0.400pt}}
\put(548,293.67){\rule{5.541pt}{0.400pt}}
\multiput(548.00,294.17)(11.500,-1.000){2}{\rule{2.770pt}{0.400pt}}
\put(503.0,295.0){\rule[-0.200pt]{10.840pt}{0.400pt}}
\put(617,292.67){\rule{5.541pt}{0.400pt}}
\multiput(617.00,293.17)(11.500,-1.000){2}{\rule{2.770pt}{0.400pt}}
\put(571.0,294.0){\rule[-0.200pt]{11.081pt}{0.400pt}}
\put(685,291.67){\rule{5.541pt}{0.400pt}}
\multiput(685.00,292.17)(11.500,-1.000){2}{\rule{2.770pt}{0.400pt}}
\put(640.0,293.0){\rule[-0.200pt]{10.840pt}{0.400pt}}
\put(753,290.67){\rule{5.541pt}{0.400pt}}
\multiput(753.00,291.17)(11.500,-1.000){2}{\rule{2.770pt}{0.400pt}}
\put(708.0,292.0){\rule[-0.200pt]{10.840pt}{0.400pt}}
\put(844,289.67){\rule{5.541pt}{0.400pt}}
\multiput(844.00,290.17)(11.500,-1.000){2}{\rule{2.770pt}{0.400pt}}
\put(776.0,291.0){\rule[-0.200pt]{16.381pt}{0.400pt}}
\put(913,288.67){\rule{5.300pt}{0.400pt}}
\multiput(913.00,289.17)(11.000,-1.000){2}{\rule{2.650pt}{0.400pt}}
\put(867.0,290.0){\rule[-0.200pt]{11.081pt}{0.400pt}}
\put(981,287.67){\rule{5.541pt}{0.400pt}}
\multiput(981.00,288.17)(11.500,-1.000){2}{\rule{2.770pt}{0.400pt}}
\put(935.0,289.0){\rule[-0.200pt]{11.081pt}{0.400pt}}
\put(1049,286.67){\rule{5.541pt}{0.400pt}}
\multiput(1049.00,287.17)(11.500,-1.000){2}{\rule{2.770pt}{0.400pt}}
\put(1004.0,288.0){\rule[-0.200pt]{10.840pt}{0.400pt}}
\put(1117,285.67){\rule{5.541pt}{0.400pt}}
\multiput(1117.00,286.17)(11.500,-1.000){2}{\rule{2.770pt}{0.400pt}}
\put(1072.0,287.0){\rule[-0.200pt]{10.840pt}{0.400pt}}
\put(1208,284.67){\rule{5.541pt}{0.400pt}}
\multiput(1208.00,285.17)(11.500,-1.000){2}{\rule{2.770pt}{0.400pt}}
\put(1140.0,286.0){\rule[-0.200pt]{16.381pt}{0.400pt}}
\put(1277,283.67){\rule{5.300pt}{0.400pt}}
\multiput(1277.00,284.17)(11.000,-1.000){2}{\rule{2.650pt}{0.400pt}}
\put(1231.0,285.0){\rule[-0.200pt]{11.081pt}{0.400pt}}
\put(1368,282.67){\rule{5.300pt}{0.400pt}}
\multiput(1368.00,283.17)(11.000,-1.000){2}{\rule{2.650pt}{0.400pt}}
\put(1299.0,284.0){\rule[-0.200pt]{16.622pt}{0.400pt}}
\put(1390.0,283.0){\rule[-0.200pt]{11.081pt}{0.400pt}}
\end{picture}
\end{center}
\caption{The ratio $R_\pi(Q^2) \equiv \frac{F_\pi (Q^2)}{4 \pi Q^2
|F_{\pi \gamma}(Q^2)|^2}$.}
\label{rpi}
\end{figure}
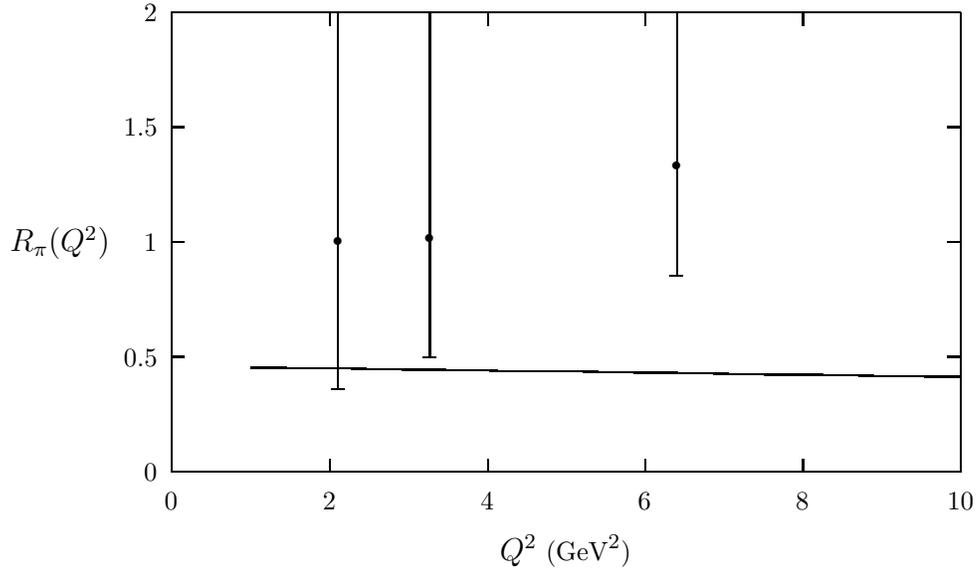

In Fig. \ref{fpi} we compare the existing measurements of the
space-like pion form factor $F_\pi(Q^2)$ \cite{PionFF1,PionFF2}
(obtained from the extrapolation of $\gamma^* p \to \pi^+ n$ data to
the pion pole) with the QCD prediction (\ref{pionformfactornlo}),
again assuming the asymptotic form of the pion distribution amplitude.
The scaling of the pion form factor data is again important evidence
for the nominal scaling of the leading twist prediction. However, the
prediction is lower than the data by approximately a factor of 2.  The
same feature can be seen in the ratio $R_\pi(Q^2)$ (Fig. \ref{rpi}),
in which the uncertainties due to the unknown form of the pion
distribution amplitude tend to cancel out.

We can estimate the sensitivity of these results to the choice of
distribution amplitude by comparing the results for the asymptotic
amplitude to, {\it e.g.}, those obtained using the Chernyak-Zhitnitshy
(CZ) form.  A full analysis at NLO of this kind is somewhat beyond the
scope of the present work, however, because of the need to include the
full ${\cal O}(\alpha_s^2)$ evolution of the CZ amplitude in order to
consistently calculate to NLO.  At LO, however, we find that $F_\pi$
is increased by roughly a factor of three for the CZ amplitude
(relative to the LO result for the asymptotic amplitude, of course),
while $F_{\gamma\pi}$ increases by a factor of about 1.5.  These
estimates are probably quite crude, but give an indication of the
typical range over which the results can vary.

\begin{figure}
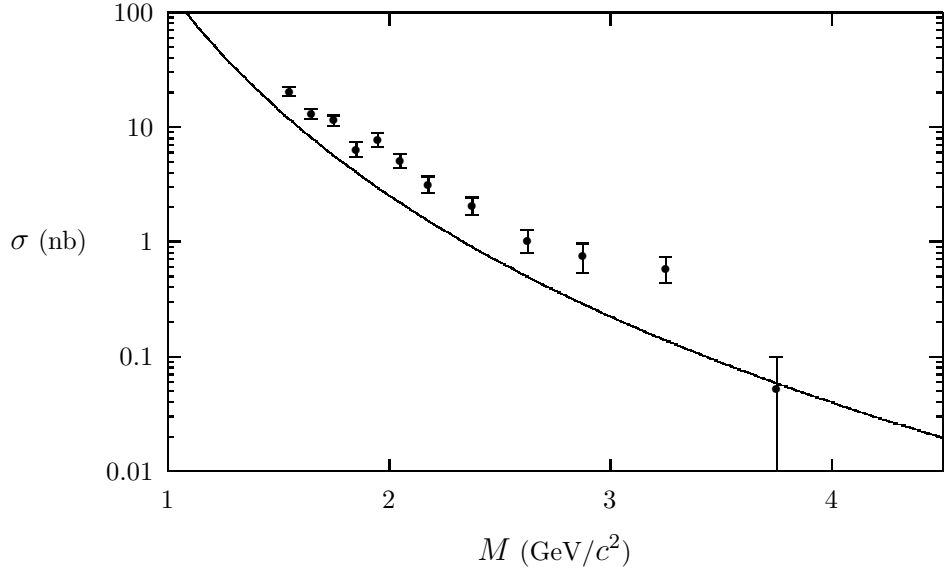

\begin{center}
\input fig6.tex
\end{center}
\caption{Two-photon annihilation cross section $\sigma(\gamma \gamma
\to\pi^+\pi^-,K^+K^-)$ as a function of CMS energy, for
$|\cos\theta^*|<0.6$.}
\label{sigma}
\end{figure}

We have also analyzed the $\gamma\gamma \rightarrow \pi^+\pi^-$ data.
These data exhibit true leading-twist scaling (Fig. \ref{sigma}), so
that one would expect this process to be a good test of theory. One
can show \cite{BL} that, to LO,
\begin{equation}
{{d\sigma\over dt}\left(\gamma\gamma\rightarrow\pi^+\pi^-\right) \over
{d\sigma\over dt}\left(\gamma\gamma\rightarrow\mu^+\mu^-\right)} =
{4|F_\pi(s)|^2\over 1-\cos^4\theta_{c.m.}}
\end{equation}
in the c.m. system (CMS), where $dt=(s/2) d(\cos\theta_{c.m.})$ and
here $F_\pi(s)$ is the {\em time-like} pion form factor.  The ratio of
the time-like to space-like pion form factor for the asymptotic
distribution amplitude is given by
\begin{equation}
{|F^{(timelike)}_\pi(-Q^2)|\over F^{(spacelike)}_\pi(Q^2)}
= {|\alpha_V(-Q^{*2})|\over \alpha_V(Q^{*2})}.
\label{ratio}
\end{equation}
If we simply continue Eq. (\ref{frozencoupling}) to negative values of
$Q^2$ (Fig. \ref{continuation}), then for $1 < Q^2 < 10$ GeV$^2$, and
hence $0.05 < Q^{*2} < 0.5$ GeV$^2$, the ratio of couplings in
Eq. (\ref{ratio}) is of order 1.5.  Of course this assumes the
analytic application of Eq. (\ref{frozencoupling}).  Thus if we assume
the asymptotic form for the distribution amplitude, then we predict
$F^{(timelike)}_\pi(-Q^2) \simeq (0.3~{\rm GeV}^2)/Q^2$ and hence
\begin{equation}
{{d\sigma\over dt}\left(\gamma\gamma\rightarrow\pi^+\pi^-\right) \over
{d\sigma\over dt}\left(\gamma\gamma\rightarrow\mu^+\mu^-\right)
}\simeq {.36 \over s^2} {1\over 1-\cos^4\theta_{c.m.}}.
\label{twophotonratio}
\end{equation}
The resulting prediction for the combined cross section\footnote{The
contribution from kaons is obtained at this order simply by rescaling
the prediction for pions by a factor $(f_K/f_\pi)^4\simeq 2.2$.}
$\sigma(\gamma\gamma\to\pi^+\pi^-, K^+K^-)$ is shown in
Fig. \ref{sigma}, along with the data of Ref. \cite{Dominick}.
Considering the possible contribution of the resonance $f_2(1270)$,
the agreement is reasonable.

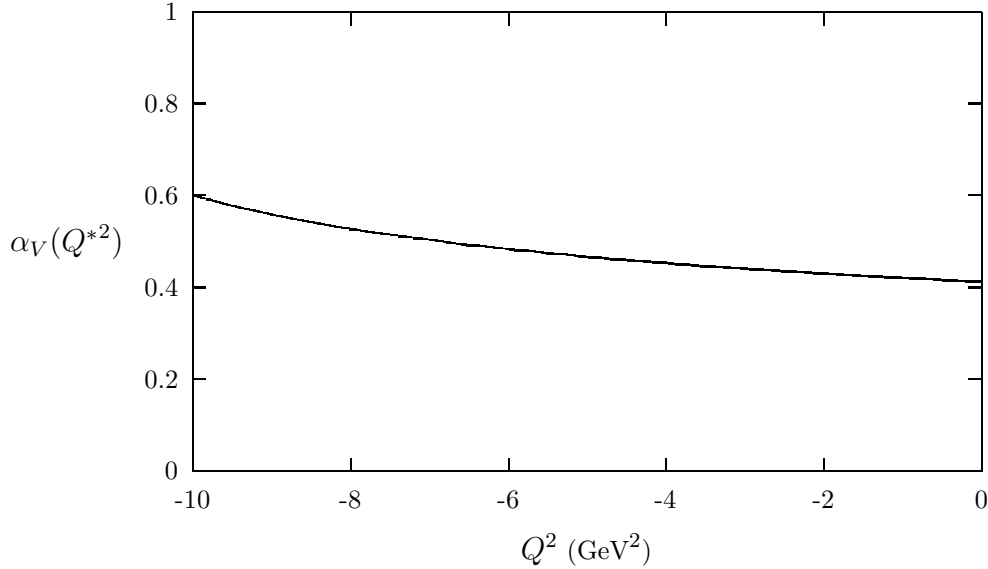
\begin{figure}
\begin{center}
\setlength{\unitlength}{0.240900pt}
\ifx\plotpoint\undefined\newsavebox{\plotpoint}\fi
\sbox{\plotpoint}{\rule[-0.200pt]{0.400pt}{0.400pt}}%
\begin{picture}(1500,900)(0,0)
\font\gnuplot=cmr10 at 10pt
\gnuplot
\sbox{\plotpoint}{\rule[-0.200pt]{0.400pt}{0.400pt}}%
\put(197.0,134.0){\rule[-0.200pt]{4.818pt}{0.400pt}}
\put(175,134){\makebox(0,0)[r]{0}}
\put(1416.0,134.0){\rule[-0.200pt]{4.818pt}{0.400pt}}
\put(197.0,278.0){\rule[-0.200pt]{4.818pt}{0.400pt}}
\put(175,278){\makebox(0,0)[r]{0.2}}
\put(1416.0,278.0){\rule[-0.200pt]{4.818pt}{0.400pt}}
\put(197.0,422.0){\rule[-0.200pt]{4.818pt}{0.400pt}}
\put(175,422){\makebox(0,0)[r]{0.4}}
\put(1416.0,422.0){\rule[-0.200pt]{4.818pt}{0.400pt}}
\put(197.0,567.0){\rule[-0.200pt]{4.818pt}{0.400pt}}
\put(175,567){\makebox(0,0)[r]{0.6}}
\put(1416.0,567.0){\rule[-0.200pt]{4.818pt}{0.400pt}}
\put(197.0,711.0){\rule[-0.200pt]{4.818pt}{0.400pt}}
\put(175,711){\makebox(0,0)[r]{0.8}}
\put(1416.0,711.0){\rule[-0.200pt]{4.818pt}{0.400pt}}
\put(197.0,855.0){\rule[-0.200pt]{4.818pt}{0.400pt}}
\put(175,855){\makebox(0,0)[r]{1}}
\put(1416.0,855.0){\rule[-0.200pt]{4.818pt}{0.400pt}}
\put(197.0,134.0){\rule[-0.200pt]{0.400pt}{4.818pt}}
\put(197,89){\makebox(0,0){-10}}
\put(197.0,835.0){\rule[-0.200pt]{0.400pt}{4.818pt}}
\put(445.0,134.0){\rule[-0.200pt]{0.400pt}{4.818pt}}
\put(445,89){\makebox(0,0){-8}}
\put(445.0,835.0){\rule[-0.200pt]{0.400pt}{4.818pt}}
\put(693.0,134.0){\rule[-0.200pt]{0.400pt}{4.818pt}}
\put(693,89){\makebox(0,0){-6}}
\put(693.0,835.0){\rule[-0.200pt]{0.400pt}{4.818pt}}
\put(940.0,134.0){\rule[-0.200pt]{0.400pt}{4.818pt}}
\put(940,89){\makebox(0,0){-4}}
\put(940.0,835.0){\rule[-0.200pt]{0.400pt}{4.818pt}}
\put(1188.0,134.0){\rule[-0.200pt]{0.400pt}{4.818pt}}
\put(1188,89){\makebox(0,0){-2}}
\put(1188.0,835.0){\rule[-0.200pt]{0.400pt}{4.818pt}}
\put(1436.0,134.0){\rule[-0.200pt]{0.400pt}{4.818pt}}
\put(1436,89){\makebox(0,0){0}}
\put(1436.0,835.0){\rule[-0.200pt]{0.400pt}{4.818pt}}
\put(197.0,134.0){\rule[-0.200pt]{298.475pt}{0.400pt}}
\put(1436.0,134.0){\rule[-0.200pt]{0.400pt}{173.689pt}}
\put(197.0,855.0){\rule[-0.200pt]{298.475pt}{0.400pt}}
\put(0,494){\makebox(0,0){$\alpha_V({Q^*}^2)$}}
\put(816,10){\makebox(0,0){$Q^2$ (GeV$^2$)}}
\put(197.0,134.0){\rule[-0.200pt]{0.400pt}{173.689pt}}
\put(197,567){\usebox{\plotpoint}}
\multiput(197.00,565.93)(1.865,-0.485){11}{\rule{1.529pt}{0.117pt}}
\multiput(197.00,566.17)(21.827,-7.000){2}{\rule{0.764pt}{0.400pt}}
\multiput(222.00,558.93)(1.942,-0.485){11}{\rule{1.586pt}{0.117pt}}
\multiput(222.00,559.17)(22.709,-7.000){2}{\rule{0.793pt}{0.400pt}}
\multiput(248.00,551.93)(2.208,-0.482){9}{\rule{1.767pt}{0.116pt}}
\multiput(248.00,552.17)(21.333,-6.000){2}{\rule{0.883pt}{0.400pt}}
\multiput(273.00,545.93)(2.714,-0.477){7}{\rule{2.100pt}{0.115pt}}
\multiput(273.00,546.17)(20.641,-5.000){2}{\rule{1.050pt}{0.400pt}}
\multiput(298.00,540.93)(2.208,-0.482){9}{\rule{1.767pt}{0.116pt}}
\multiput(298.00,541.17)(21.333,-6.000){2}{\rule{0.883pt}{0.400pt}}
\multiput(323.00,534.93)(2.825,-0.477){7}{\rule{2.180pt}{0.115pt}}
\multiput(323.00,535.17)(21.475,-5.000){2}{\rule{1.090pt}{0.400pt}}
\multiput(349.00,529.93)(2.714,-0.477){7}{\rule{2.100pt}{0.115pt}}
\multiput(349.00,530.17)(20.641,-5.000){2}{\rule{1.050pt}{0.400pt}}
\multiput(374.00,524.94)(3.552,-0.468){5}{\rule{2.600pt}{0.113pt}}
\multiput(374.00,525.17)(19.604,-4.000){2}{\rule{1.300pt}{0.400pt}}
\multiput(399.00,520.93)(2.825,-0.477){7}{\rule{2.180pt}{0.115pt}}
\multiput(399.00,521.17)(21.475,-5.000){2}{\rule{1.090pt}{0.400pt}}
\multiput(425.00,515.94)(3.552,-0.468){5}{\rule{2.600pt}{0.113pt}}
\multiput(425.00,516.17)(19.604,-4.000){2}{\rule{1.300pt}{0.400pt}}
\multiput(450.00,511.94)(3.552,-0.468){5}{\rule{2.600pt}{0.113pt}}
\multiput(450.00,512.17)(19.604,-4.000){2}{\rule{1.300pt}{0.400pt}}
\multiput(475.00,507.95)(5.374,-0.447){3}{\rule{3.433pt}{0.108pt}}
\multiput(475.00,508.17)(17.874,-3.000){2}{\rule{1.717pt}{0.400pt}}
\multiput(500.00,504.94)(3.698,-0.468){5}{\rule{2.700pt}{0.113pt}}
\multiput(500.00,505.17)(20.396,-4.000){2}{\rule{1.350pt}{0.400pt}}
\multiput(526.00,500.95)(5.374,-0.447){3}{\rule{3.433pt}{0.108pt}}
\multiput(526.00,501.17)(17.874,-3.000){2}{\rule{1.717pt}{0.400pt}}
\multiput(551.00,497.95)(5.374,-0.447){3}{\rule{3.433pt}{0.108pt}}
\multiput(551.00,498.17)(17.874,-3.000){2}{\rule{1.717pt}{0.400pt}}
\multiput(576.00,494.94)(3.698,-0.468){5}{\rule{2.700pt}{0.113pt}}
\multiput(576.00,495.17)(20.396,-4.000){2}{\rule{1.350pt}{0.400pt}}
\multiput(602.00,490.95)(5.374,-0.447){3}{\rule{3.433pt}{0.108pt}}
\multiput(602.00,491.17)(17.874,-3.000){2}{\rule{1.717pt}{0.400pt}}
\put(627,487.17){\rule{5.100pt}{0.400pt}}
\multiput(627.00,488.17)(14.415,-2.000){2}{\rule{2.550pt}{0.400pt}}
\multiput(652.00,485.95)(5.374,-0.447){3}{\rule{3.433pt}{0.108pt}}
\multiput(652.00,486.17)(17.874,-3.000){2}{\rule{1.717pt}{0.400pt}}
\multiput(677.00,482.95)(5.597,-0.447){3}{\rule{3.567pt}{0.108pt}}
\multiput(677.00,483.17)(18.597,-3.000){2}{\rule{1.783pt}{0.400pt}}
\put(703,479.17){\rule{5.100pt}{0.400pt}}
\multiput(703.00,480.17)(14.415,-2.000){2}{\rule{2.550pt}{0.400pt}}
\multiput(728.00,477.95)(5.374,-0.447){3}{\rule{3.433pt}{0.108pt}}
\multiput(728.00,478.17)(17.874,-3.000){2}{\rule{1.717pt}{0.400pt}}
\put(753,474.17){\rule{5.300pt}{0.400pt}}
\multiput(753.00,475.17)(15.000,-2.000){2}{\rule{2.650pt}{0.400pt}}
\multiput(779.00,472.95)(5.374,-0.447){3}{\rule{3.433pt}{0.108pt}}
\multiput(779.00,473.17)(17.874,-3.000){2}{\rule{1.717pt}{0.400pt}}
\put(804,469.17){\rule{5.100pt}{0.400pt}}
\multiput(804.00,470.17)(14.415,-2.000){2}{\rule{2.550pt}{0.400pt}}
\put(829,467.17){\rule{5.100pt}{0.400pt}}
\multiput(829.00,468.17)(14.415,-2.000){2}{\rule{2.550pt}{0.400pt}}
\put(854,465.17){\rule{5.300pt}{0.400pt}}
\multiput(854.00,466.17)(15.000,-2.000){2}{\rule{2.650pt}{0.400pt}}
\put(880,463.17){\rule{5.100pt}{0.400pt}}
\multiput(880.00,464.17)(14.415,-2.000){2}{\rule{2.550pt}{0.400pt}}
\put(905,461.17){\rule{5.100pt}{0.400pt}}
\multiput(905.00,462.17)(14.415,-2.000){2}{\rule{2.550pt}{0.400pt}}
\put(930,459.17){\rule{5.300pt}{0.400pt}}
\multiput(930.00,460.17)(15.000,-2.000){2}{\rule{2.650pt}{0.400pt}}
\put(956,457.17){\rule{5.100pt}{0.400pt}}
\multiput(956.00,458.17)(14.415,-2.000){2}{\rule{2.550pt}{0.400pt}}
\put(981,455.17){\rule{5.100pt}{0.400pt}}
\multiput(981.00,456.17)(14.415,-2.000){2}{\rule{2.550pt}{0.400pt}}
\put(1006,453.67){\rule{6.023pt}{0.400pt}}
\multiput(1006.00,454.17)(12.500,-1.000){2}{\rule{3.011pt}{0.400pt}}
\put(1031,452.17){\rule{5.300pt}{0.400pt}}
\multiput(1031.00,453.17)(15.000,-2.000){2}{\rule{2.650pt}{0.400pt}}
\put(1057,450.17){\rule{5.100pt}{0.400pt}}
\multiput(1057.00,451.17)(14.415,-2.000){2}{\rule{2.550pt}{0.400pt}}
\put(1082,448.67){\rule{6.023pt}{0.400pt}}
\multiput(1082.00,449.17)(12.500,-1.000){2}{\rule{3.011pt}{0.400pt}}
\put(1107,447.17){\rule{5.300pt}{0.400pt}}
\multiput(1107.00,448.17)(15.000,-2.000){2}{\rule{2.650pt}{0.400pt}}
\put(1133,445.67){\rule{6.023pt}{0.400pt}}
\multiput(1133.00,446.17)(12.500,-1.000){2}{\rule{3.011pt}{0.400pt}}
\put(1158,444.17){\rule{5.100pt}{0.400pt}}
\multiput(1158.00,445.17)(14.415,-2.000){2}{\rule{2.550pt}{0.400pt}}
\put(1183,442.67){\rule{6.023pt}{0.400pt}}
\multiput(1183.00,443.17)(12.500,-1.000){2}{\rule{3.011pt}{0.400pt}}
\put(1208,441.17){\rule{5.300pt}{0.400pt}}
\multiput(1208.00,442.17)(15.000,-2.000){2}{\rule{2.650pt}{0.400pt}}
\put(1234,439.67){\rule{6.023pt}{0.400pt}}
\multiput(1234.00,440.17)(12.500,-1.000){2}{\rule{3.011pt}{0.400pt}}
\put(1259,438.17){\rule{5.100pt}{0.400pt}}
\multiput(1259.00,439.17)(14.415,-2.000){2}{\rule{2.550pt}{0.400pt}}
\put(1284,436.67){\rule{6.263pt}{0.400pt}}
\multiput(1284.00,437.17)(13.000,-1.000){2}{\rule{3.132pt}{0.400pt}}
\put(1310,435.67){\rule{6.023pt}{0.400pt}}
\multiput(1310.00,436.17)(12.500,-1.000){2}{\rule{3.011pt}{0.400pt}}
\put(1335,434.67){\rule{6.023pt}{0.400pt}}
\multiput(1335.00,435.17)(12.500,-1.000){2}{\rule{3.011pt}{0.400pt}}
\put(1360,433.17){\rule{5.100pt}{0.400pt}}
\multiput(1360.00,434.17)(14.415,-2.000){2}{\rule{2.550pt}{0.400pt}}
\put(1385,431.67){\rule{6.263pt}{0.400pt}}
\multiput(1385.00,432.17)(13.000,-1.000){2}{\rule{3.132pt}{0.400pt}}
\put(1411,430.67){\rule{6.023pt}{0.400pt}}
\multiput(1411.00,431.17)(12.500,-1.000){2}{\rule{3.011pt}{0.400pt}}
\end{picture}
\end{center}
\caption{Continuation of Eq. (\ref{frozencoupling}) to negative
$Q^2$.  Note that ${Q^*}^2 \equiv e^{-3}Q^2$.}
\label{continuation}
\end{figure}

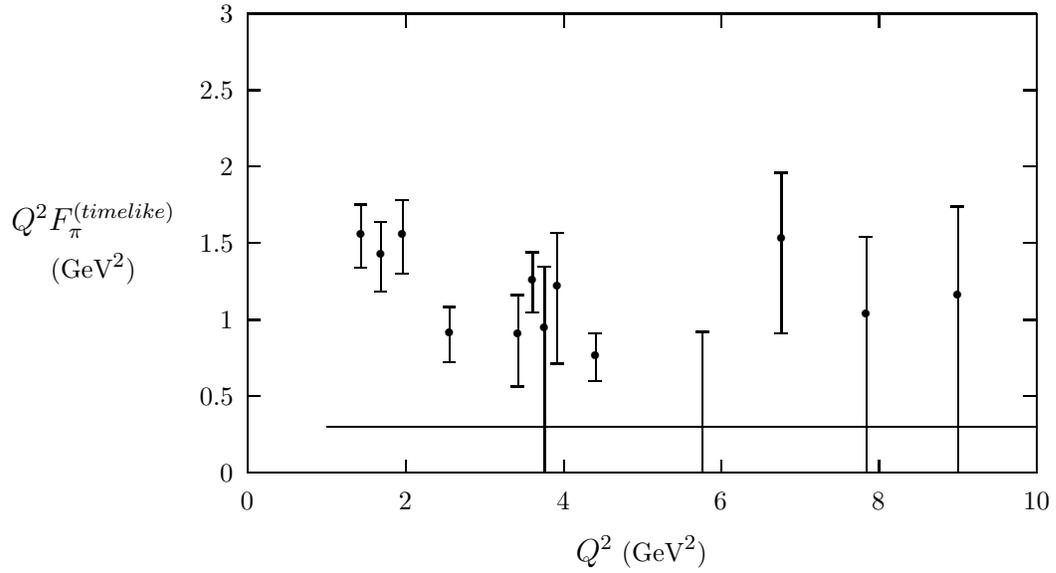
\begin{figure}
\begin{center}
\setlength{\unitlength}{0.240900pt}
\ifx\plotpoint\undefined\newsavebox{\plotpoint}\fi
\sbox{\plotpoint}{\rule[-0.200pt]{0.400pt}{0.400pt}}%
\begin{picture}(1500,900)(0,0)
\font\gnuplot=cmr10 at 10pt
\gnuplot
\sbox{\plotpoint}{\rule[-0.200pt]{0.400pt}{0.400pt}}%
\put(197.0,134.0){\rule[-0.200pt]{4.818pt}{0.400pt}}
\put(175,134){\makebox(0,0)[r]{0}}
\put(1416.0,134.0){\rule[-0.200pt]{4.818pt}{0.400pt}}
\put(197.0,254.0){\rule[-0.200pt]{4.818pt}{0.400pt}}
\put(175,254){\makebox(0,0)[r]{0.5}}
\put(1416.0,254.0){\rule[-0.200pt]{4.818pt}{0.400pt}}
\put(197.0,374.0){\rule[-0.200pt]{4.818pt}{0.400pt}}
\put(175,374){\makebox(0,0)[r]{1}}
\put(1416.0,374.0){\rule[-0.200pt]{4.818pt}{0.400pt}}
\put(197.0,495.0){\rule[-0.200pt]{4.818pt}{0.400pt}}
\put(175,495){\makebox(0,0)[r]{1.5}}
\put(1416.0,495.0){\rule[-0.200pt]{4.818pt}{0.400pt}}
\put(197.0,615.0){\rule[-0.200pt]{4.818pt}{0.400pt}}
\put(175,615){\makebox(0,0)[r]{2}}
\put(1416.0,615.0){\rule[-0.200pt]{4.818pt}{0.400pt}}
\put(197.0,735.0){\rule[-0.200pt]{4.818pt}{0.400pt}}
\put(175,735){\makebox(0,0)[r]{2.5}}
\put(1416.0,735.0){\rule[-0.200pt]{4.818pt}{0.400pt}}
\put(197.0,855.0){\rule[-0.200pt]{4.818pt}{0.400pt}}
\put(175,855){\makebox(0,0)[r]{3}}
\put(1416.0,855.0){\rule[-0.200pt]{4.818pt}{0.400pt}}
\put(197.0,134.0){\rule[-0.200pt]{0.400pt}{4.818pt}}
\put(197,89){\makebox(0,0){0}}
\put(197.0,835.0){\rule[-0.200pt]{0.400pt}{4.818pt}}
\put(445.0,134.0){\rule[-0.200pt]{0.400pt}{4.818pt}}
\put(445,89){\makebox(0,0){2}}
\put(445.0,835.0){\rule[-0.200pt]{0.400pt}{4.818pt}}
\put(693.0,134.0){\rule[-0.200pt]{0.400pt}{4.818pt}}
\put(693,89){\makebox(0,0){4}}
\put(693.0,835.0){\rule[-0.200pt]{0.400pt}{4.818pt}}
\put(940.0,134.0){\rule[-0.200pt]{0.400pt}{4.818pt}}
\put(940,89){\makebox(0,0){6}}
\put(940.0,835.0){\rule[-0.200pt]{0.400pt}{4.818pt}}
\put(1188.0,134.0){\rule[-0.200pt]{0.400pt}{4.818pt}}
\put(1188,89){\makebox(0,0){8}}
\put(1188.0,835.0){\rule[-0.200pt]{0.400pt}{4.818pt}}
\put(1436.0,134.0){\rule[-0.200pt]{0.400pt}{4.818pt}}
\put(1436,89){\makebox(0,0){10}}
\put(1436.0,835.0){\rule[-0.200pt]{0.400pt}{4.818pt}}
\put(197.0,134.0){\rule[-0.200pt]{298.475pt}{0.400pt}}
\put(1436.0,134.0){\rule[-0.200pt]{0.400pt}{173.689pt}}
\put(197.0,855.0){\rule[-0.200pt]{298.475pt}{0.400pt}}
\put(-45,494){\makebox(0,0){\shortstack{$Q^2 F_\pi^{(timelike)}$\\ \\ (GeV$^2$)}}}
\put(816,10){\makebox(0,0){$Q^2$ (GeV$^2$)}}
\put(197.0,134.0){\rule[-0.200pt]{0.400pt}{173.689pt}}
\put(375,509){\circle*{12}}
\put(406,478){\circle*{12}}
\put(440,509){\circle*{12}}
\put(514,355){\circle*{12}}
\put(621,353){\circle*{12}}
\put(644,437){\circle*{12}}
\put(663,362){\circle*{12}}
\put(683,427){\circle*{12}}
\put(743,319){\circle*{12}}
\put(1035,502){\circle*{12}}
\put(1168,384){\circle*{12}}
\put(1312,413){\circle*{12}}
\put(375.0,456.0){\rule[-0.200pt]{0.400pt}{23.849pt}}
\put(365.0,456.0){\rule[-0.200pt]{4.818pt}{0.400pt}}
\put(365.0,555.0){\rule[-0.200pt]{4.818pt}{0.400pt}}
\put(406.0,418.0){\rule[-0.200pt]{0.400pt}{26.499pt}}
\put(396.0,418.0){\rule[-0.200pt]{4.818pt}{0.400pt}}
\put(396.0,528.0){\rule[-0.200pt]{4.818pt}{0.400pt}}
\put(440.0,446.0){\rule[-0.200pt]{0.400pt}{27.944pt}}
\put(430.0,446.0){\rule[-0.200pt]{4.818pt}{0.400pt}}
\put(430.0,562.0){\rule[-0.200pt]{4.818pt}{0.400pt}}
\put(514.0,307.0){\rule[-0.200pt]{0.400pt}{20.958pt}}
\put(504.0,307.0){\rule[-0.200pt]{4.818pt}{0.400pt}}
\put(504.0,394.0){\rule[-0.200pt]{4.818pt}{0.400pt}}
\put(621.0,269.0){\rule[-0.200pt]{0.400pt}{34.690pt}}
\put(611.0,269.0){\rule[-0.200pt]{4.818pt}{0.400pt}}
\put(611.0,413.0){\rule[-0.200pt]{4.818pt}{0.400pt}}
\put(644.0,386.0){\rule[-0.200pt]{0.400pt}{22.645pt}}
\put(634.0,386.0){\rule[-0.200pt]{4.818pt}{0.400pt}}
\put(634.0,480.0){\rule[-0.200pt]{4.818pt}{0.400pt}}
\put(663.0,134.0){\rule[-0.200pt]{0.400pt}{78.052pt}}
\put(653.0,134.0){\rule[-0.200pt]{4.818pt}{0.400pt}}
\put(653.0,458.0){\rule[-0.200pt]{4.818pt}{0.400pt}}
\put(683.0,305.0){\rule[-0.200pt]{0.400pt}{49.625pt}}
\put(673.0,305.0){\rule[-0.200pt]{4.818pt}{0.400pt}}
\put(673.0,511.0){\rule[-0.200pt]{4.818pt}{0.400pt}}
\put(743.0,278.0){\rule[-0.200pt]{0.400pt}{18.067pt}}
\put(733.0,278.0){\rule[-0.200pt]{4.818pt}{0.400pt}}
\put(733.0,353.0){\rule[-0.200pt]{4.818pt}{0.400pt}}
\put(911.0,134.0){\rule[-0.200pt]{0.400pt}{53.239pt}}
\put(901.0,134.0){\rule[-0.200pt]{4.818pt}{0.400pt}}
\put(901.0,355.0){\rule[-0.200pt]{4.818pt}{0.400pt}}
\put(1035.0,353.0){\rule[-0.200pt]{0.400pt}{60.707pt}}
\put(1025.0,353.0){\rule[-0.200pt]{4.818pt}{0.400pt}}
\put(1025.0,605.0){\rule[-0.200pt]{4.818pt}{0.400pt}}
\put(1168.0,134.0){\rule[-0.200pt]{0.400pt}{89.133pt}}
\put(1158.0,134.0){\rule[-0.200pt]{4.818pt}{0.400pt}}
\put(1158.0,504.0){\rule[-0.200pt]{4.818pt}{0.400pt}}
\put(1312.0,134.0){\rule[-0.200pt]{0.400pt}{100.696pt}}
\put(1302.0,134.0){\rule[-0.200pt]{4.818pt}{0.400pt}}
\put(1302.0,552.0){\rule[-0.200pt]{4.818pt}{0.400pt}}
\put(321,206){\usebox{\plotpoint}}
\put(321.0,206.0){\rule[-0.200pt]{268.603pt}{0.400pt}}
\end{picture}
\end{center}
\caption{Pion electromagnetic form factor in the time-like region.}
\label{fpi-timelike}
\end{figure}

It should be noted that the leading-twist prediction $Q^2
F^{(timelike)}_\pi (-Q^2) = 0.3$ GeV$^2$ is a factor of two below the
measurement of the pion form factor obtained from the
$J/\psi\rightarrow\pi^+\pi^-$ branching ratio. The $J/\psi$ analysis
assumes that the $\pi^+\pi^-$ is created only through virtual photons.
However, if the $J/\psi\rightarrow\pi^+\pi^-$ amplitude proceeds
through channels such as $\gamma gg$, then the branching ratio is not
a precise method for obtaining $F^{(timelike)}_\pi$.  It is thus
important to have direct measurement of the $e^+e^-\to\pi^+\pi^-$
amplitude off-resonance.  We also show the prediction for the pion
form factor in the time-like region compared with the data of Bollini
{\it et al.} \cite{Bollini} in Fig. \ref{fpi-timelike}.  We emphasize
that the normalization of the prediction
\begin{eqnarray}
F^{(timelike)}_\pi(-Q^2)&=&{16\pi f^2_\pi\over Q^2}
\alpha_V(-{Q^*}^2)\left(1-1.9{\alpha_V\over\pi}\right) \\
& \simeq & {0.3~{\rm GeV}^2 \over Q^2} \nonumber
\end{eqnarray}
assumes the asymptotic form for the pion distribution amplitude and
the form of $\alpha_V$ given in Eq. (\ref{frozencoupling}), with the
parameters $m^2_g=0.19$ GeV$^2$ and $\Lambda_V=0.16$ GeV.  There is
clearly some room to readjust these parameters.  However, even at the
initial stage of approximation done in this paper, which includes NLO
corrections at the BLM scale, there is no significant discrepancy with
the relevant experiments.

The values for the space-like pion form factor $F_\pi(Q^2)$ obtained
from the extrapolation of $\gamma^* p \to \pi^+ n$ data to the pion
pole thus appear to be systematically higher in normalization than
predicted by commensurate scale relations; however, it should be
emphasized that this discrepancy may be due to systematic errors
introduced by the extrapolation procedure \cite{CarlsonMilana}. What
is at best measured in electroproduction is the transition amplitude
between a mesonic state with an effective space-like mass $m^2=t < 0$
and the physical pion. It is theoretically possible that the off-shell
form factor $F_\pi(Q^2,t)$ is significantly larger than the physical
form factor because of its bias towards more point-like $q \overline
q$ valence configurations in its Fock state structure.  The
extrapolation to the pole at $t=m^2_\pi$ also requires knowing the
analytic dependence of $F_\pi(Q^2,t)$ on $t$.  These considerations
are discussed further in Ref. \cite{BR}.  If we assume that there are
no significant errors induced by the electroproduction extrapolation,
then one must look for other sources for the discrepancy in
normalization.  Note that the NLO corrections in
Eqs. (\ref{pionformfactornlo}) and (\ref{RpiV}) are of order 20--30\%.
Thus there may be large contributions from NNLO and higher corrections
which need to be re-summed. There are also possible corrections from
pion rescattering in the final state of the electroproduction process.
It thus would be very interesting to have unambiguous data on the pion
form factors from electron-pion collisions, say, by scattering
electrons on a secondary pion beam at the SLAC Linear Collider.  In
addition, it is possible that pre-asymptotic contributions from
higher-twist or soft Feynman-type physics are important.

We also note that the normalization of $\alpha_V$ could be larger at
low momentum than our estimate. This would also imply a broadening of
the pion distribution amplitude compared to its asymptotic form since
one needs to raise the expectation value of $1/(1-x)$ in order to
maintain consistency with the magnitude of the $F_{\gamma \pi}(Q^2)$
data.  A full analysis will then also require consideration of the
breaking of scaling from the evolution of the distribution amplitude.

In any case, we find no compelling argument for significant
higher-twist contributions in the few GeV regime from the hard
scattering amplitude or the endpoint regions, since such corrections
violate the observed scaling behavior of the data.

The time-like pion form factor data obtained from $e^+ e^- \to \pi^+
\pi^-$ annihilation does not have complications from off-shell
extrapolations or rescattering, but it is also more sensitive to
nearby vector meson poles in the $t$ channel. If we analytically
continue the leading twist prediction and the effective form of
$\alpha_V$ to the time-like regime, we obtain the prediction shown in
Fig. \ref{fpi-timelike}, again assuming the asymptotic form of the
pion distribution amplitude.

The analysis we have presented here suggests a systematic program for
estimating exclusive amplitudes in QCD.  The central input is
$\alpha_V(0)$, or
\begin{equation}
\overline{\alpha_V} = {1\over{Q_0^2}}\int_0^{Q_0^2}d{Q^\prime}^2
\alpha_V({Q^\prime}^2),\;\; Q_0^2 \leq 1\;\;{\rm GeV}^2,
\label{old21}
\end{equation}
which largely controls the magnitude of the underlying quark-gluon
subprocesses for hard processes in the few-GeV region.  In this work,
the mean coupling value for $Q_0^2 \simeq 0.5$ GeV$^2$ is
$\overline{\alpha_V} \simeq 0.38.$ The main focus will then be to
determine the shapes and normalization of the process-independent
meson and baryon distribution amplitudes.

\section{Conclusions}

In this paper we have shown that dimensional counting rules emerge if
the effective coupling $\alpha_V(Q^*)$ is approximately constant in
the domain of $Q^*$ relevant to the hard scattering amplitudes of
exclusive processes. In the low-$Q^*$ domain, evolution of the quark
distribution amplitudes is also minimal. Furthermore, Sudakov
suppression of the long-distance contributions is strengthened if the
coupling is frozen because of the exponentiation of a double log
series.  The Ansatz of a frozen coupling at small momentum transfer
has not been demonstrated from first principles. However, the behavior
of exclusive amplitudes point strongly to scaling behavior in the
kinematic regions we discussed.  We have also found that the CSR
connecting the heavy quark potential, as determined from lattice gauge
theory, to the photon-to-pion transition form factor is in excellent
agreement with $\gamma e \to \pi^0 e$ data assuming that the pion
distribution amplitude is close to its asymptotic form $\sqrt{3}f_\pi
x(1-x)$.  We also reproduce the scaling and normalization of the
$\gamma\gamma \rightarrow \pi^+\pi^-$ data at large momentum transfer.
However, the normalization of the space-like pion form factor
$F_\pi(Q^2)$ obtained from electroproduction experiments is somewhat
higher than that predicted by the corresponding commensurate scale
relation. This discrepancy may be due to systematic errors introduced
by the extrapolation of the $\gamma^* p \to \pi^+ n$ electroproduction
data to the pion pole.

\vfill\pagebreak
\centerline{\bf Acknowledgements}

\noindent
It is a pleasure to thank Hung Jung Lu for many valuable discussions
during the early stages of this work.
We also thank
Martin Beneke and Volodya Braun for helpful conversations.
S.J.B. is supported in part by the U.S. Department of Energy under
contract no. DE--AC03--76SF00515.
C.-R.J. is supported in part by the U.S. Department of Energy under
contract no. DE--FG02--96ER40947.  The North Carolina Supercomputer
Center is also acknowledged for the grant of supercomputing time
allocation.
A.P. is supported in part by the National Science Foundation under
research contract NSF--PHY94--08843.
D.G.R. is supported in part by the U.S. Department of Energy under
contract no. DE--FG02--91ER40690.

\vfill\pagebreak

\end{document}